\documentclass[amssymb, nofootinbib, superscriptaddress]{revtex4}

\usepackage{amsmath}
\usepackage{amsfonts}
\usepackage{graphicx}
\usepackage{psfrag}
\usepackage{array}
\usepackage{bbm}
\usepackage{hyperref}
\usepackage{amsthm}
\theoremstyle{definition}

\newcommand{\f}{\frac}

\newcommand{\R}{\mathbb{R}}
\newcommand{\C}{\mathbb{C}}

\newcommand{\Z}{\mathbb{Z}}

\newcommand{\be}{\begin{equation}}
\newcommand{\ee}{\end{equation}}
\newcommand{\bes}{\begin{eqnarray}}
\newcommand{\ees}{\end{eqnarray}}
\newcommand{\mone}{^{-1}}

\def\dia{\diamond}
\def\st{\divideontimes}
\def\ov{\overline}

\def\f{\frac}  
\def\dag{^\dagger}
\def\tl{\widetilde}

\def\ka{\kappa}
\def\cc{{\cal C}}

\def\com#1{[ #1 ]}
\def\act{\rhd}
\def\mn{{\mu\nu}}

\def\demi{\frac{1}{2} }
\def\ie{{i.e. \/}}

\def\dr{{\rightarrow}}
\newcommand{\one}{\mbox{$1 \hspace{-1.0mm}  {\bf l}$}}

\def\aa{{\cal A}}  \def\cc{{\cal C}} \def\dd{{\cal D}}   \def\ggg{{\cal G}}
\def\hh{{\cal H}}   \def\kk{{\cal K}} \def\Ll{{\cal L}}  \def\nn{{\nonumber}}
    \def\ss{{\cal S}}

\def\hphi{{\hat \phi}}
\def\hpsi{{\hat \psi}}

\def\act{{\, \triangleright\, }}

\newcommand{\so}{\mathfrak{so}}
\newcommand{\SU}{\mathrm{SU}}
\newcommand{\AN}{\mathrm{AN}}

\newcommand{\SO}{\mathrm{SO}}
\newcommand{\SL}{\mathrm{SL}}
\newcommand{\sll}{\mathrm{sl}}

\def\kkk{{\mathfrak{k}}}
\def\qqq{{\mathfrak{r}}}
\def\lll{{\mathfrak{q}}}
\def\oll{{\mathfrak{b}}}

\begin{document}
%
%**************
%* Front Page *
%**************
\title{\Large\bf Scalar field theory in
 Snyder space-time: alternatives}

\author{Florian Girelli}\email{girelli@physics.usyd.edu.au}
\affiliation{School of Physics, University of Sydney, Sydney, New South Wales 2006, Australia}

\author{Etera R. Livine}\email{etera.livine@ens-lyon.fr}
\affiliation{Laboratoire de Physique, ENS Lyon, CNRS UMR 5672, 46 All\'ee d'Italie, 69007 Lyon, France}

\date{\small \today}

%\maketitle

\begin{abstract}
We construct two types of scalar field theory on Snyder space-time. The first one is based on the natural momenta addition inherent to the coset momentum space. This construction uncovers  a non-associative deformation of the Poincar\'e symmetries. 
The second one considers Snyder space-time as a subspace of a larger non-commutative space. We discuss different possibilities to restrict the extra-dimensional scalar field theory to a  theory living only on Sndyer space-time and present the consequences of these restrictions on  the Poincar\'e symmetries.  We show moreover  how the non-associative approach and the Doplicher-Fredenhagen-Roberts space can be seen as specific approximations of the extra-dimensional theory.  These results are obtained for the  3d Euclidian Snyder space-time constructed from $\SO(3,1)/\SO(3)$, but our results extend to any dimension and signature.  
\end{abstract}

\maketitle
%%%%%%%%%%%%%%%%%%%%%%%%%%%%%%%%%%%%%%%%%%%%%%%%%%%%%
\tableofcontents
\newpage
%%%%%%%%%%
\section*{Introduction}
\label{intro}
%%%%%%%%%%

After the inception of quantum field theory, it was quickly understood that ultraviolet (UV) divergencies were plaguing the theory. To introduce a UV cutoff was the natural solution to this problem, so  quantum  physicists at that time wondered about the implementation of a UV cutoff or equivalently of a minimum length, which would be consistent with the Lorentz symmetries \cite{werner}. Snyder managed to construct in 1947 a space where there is a minimum length, while at the same time the Lorentz symmetry is preserved \cite{snyder}. However, he did not construct  the (quantum) field theory on this space to check his proposal was achieving its initial motivation.

Snyder space-time consists in defining non-commutative coordinates operators $X_\mu$ which can be realized as the Lie algebra generators
$$X_\mu\sim\frac{1}{\kappa}J_{4\mu} \in \so(4,1)/\so(3,1).$$
$\ka$ is  identified with the Planck mass ($\hbar= c=1$).  The space coordinates are then represented as operators encoding rotations and have therefore a discrete spectrum: space acquires a discrete structure. The coordinates satisfy then   Snyder's commutation relations
\be\label{snyder commutation} [X_\mu,X_\nu]=i\frac{1}{\ka^2}J_{\mn},\ee
where $J_\mn$ is an infinitesimal Lorentz transformation. Momentum space is  identified as the  de Sitter space  $dS\sim\SO(4,1)/\SO(3,1)$  and Snyder picked up specific coordinates $P_\mu$ to encode momentum
$$P_\mu = \ka \frac{v_\mu}{v_4} \textrm{ with } v_A\in \R^5 \textrm{ and } -v_0^2+ v_i^2 + v_4^2=1.$$
Snyder did not discuss the choice of momenta addition to describe the many particles states. Different generalizations of Snyder's space can be considered.  For instance, one can consider the Cartan decomposition $SO(4,1)\sim \hh_4\,.\, \SO(4)$, where $\hh_4$ is isomorphic to the two sheeted 4d hyperboloid, and the decomposition $\SO(5)\sim S^4 \,.\,  \SO(4)$, where $S^4$ is isomorphic to the 4d sphere.  Snyder spaces associated to the momentum spaces $\hh_4$ and $S^4$ are Euclidian, that is there is the metric on Snyder's momentum space has signature $+,..,+$. This construction can be extended  to any dimensions and signatures: we consider the group $G=\SO(p,q)$ and the subgroup $H=\SO(p-1,q)$, then momentum space will be given by the homogeneous space $G/H$ and Snyder's  coordinates are Lie algebra generators of $\so(p,q)/\so(p-1,q)$.

The renormalization process for quantum field theory was understood during the 50's and  Snyder space-time was  forgotten until non-commutative
geometry re-appeared as a topic of interest per se. Snyder space-time is now  commonly presented as an example of non-commutative geometry which regulates the UV divergencies due to the presence of the minimal length (different field theories in Snyder space-time have been considered  see \cite{othersnyder} and references therein). It is also often brought as a possible example of effective space-time to describe quantum gravity phenomenology \cite{gol}, just as the $\kappa$-Minkowski space-time in Deformed Special Relativity \cite{amel}. The latter is a non-commutative geometry where momentum space is also the de Sitter space (obtained from the Iwasawa decomposition of $\SO(4,1)$ \cite{iwasawa, majid, dsr-group}), but the non-commutativity is of the Lie algebra type, that is $[X_\mu,X_\nu]=\frac{1}{\ka} C_\mn ^\alpha \, X_\alpha$. A lot of works has been devoted to this type of non-commutative spaces, such as determining different star products \cite{kmink to mink, star}, the differential calculus \cite{sitarz, battista}, the conserved charges \cite{laurent-kowalski, amelino}. One of the key features of these non-commutative spaces is that momentum space has the structure of a (non-abelian) Lie group. The group law does encode the notion of addition of momenta, which is then non-commutative (but still associative). Thanks to this addition, it is then straightforward to introduce a generalized Fourier transform and a  star product between plane-waves which encodes the non-abelian momenta addition. The deformation of the momenta addition can  be related to a deformation of the action of translations (the Lorentz symmetries can also be deformed), which have been classified in \cite{zak}. We would like here to export these ideas to the Snyder case, where momentum space is not a group and is obtained from the decompositions discussed above, to construct a scalar field theory living on Snyder space-time.

\medskip

Since Snyder has not specified the type of momenta addition, we shall explore  two natural possibilities. The first is obtained from the product on the coset space\footnote{We have presented a shorter version of our results in the proceedings of the Planck scale conference \cite{girliv}. We illustrated the construction for the 3d Euclidian Snyder space to perform  simple computations. The recent  work \cite{battisti} considered a different route to obtain similar results in the Lorentzian 4d case, at first order.}, whereas the second one is obtained from the group in which momentum space is naturally embedded. The key difference between these two choices is the associativity  property. Indeed, the product on the coset is obviously non-associative while the product on the group is naturally associative but involves extra degrees of freedom (the ``Lorentz sector") which need to be integrated over later.  In both cases, we define a Fourier transform and show how to recover a non-commutative space-time defined in terms of a star product which describes Snyder commutation relation \eqref{snyder commutation}. The physical interpretation of this commutation relation is different in each approach. In the non-associative approach, the Lorentz sector $J_\mn$ has no direct physical interpretation, whereas in the associative case this sector requires  extra momentum coordinates which can be considered as extra-dimensions. Actually this perspective has been already considered in the Doplicher-Fredenhagen-Roberts non-commutative space, which shares many features with the associative version of Snyder space-time. In fact we shall recall how these spaces can be related through a specific limit.

In the following we shall focus on the specific example $G=\SO(3,1)^+$, the group of  orthochronous Lorentz transformations  and its subgroup $H=\SO(3)$.  We are therefore considering the Cartan decomposition of $\SO(3,1)^+$ and we  will deal with the Euclidian 3d Snyder space. Momentum space will be given by the upper hyperboloid $\hh$. This choice enables us to perform simple calculations and provides a nice laboratory to illustrate the main ideas. Moreover,  the group $\SO(3)$ being compact, it will allow the definition of models which could not be defined if dealing with a non-compact subgroup.     All calculations can  be extended to higher dimensions and other signature in a straightforward manner.

In the first section, we discuss the momenta addition inherited from the group structure of $\SO(3,1)^+$, interpreted as a momentum space. We give in particular the explicit shape of the non-associative momenta addition obtained from the coset structure $\SO(3,1)^+/\SO(3)$.

In the second section we present a scalar field theory defined over the (upper) hyperboloid interpreted as momentum space. We define in particular the dual non-commutative space through a Fourier transform and a star product. We show how this latter encodes  Snyder's commutation relations. The key feature of this star product is its non-associativity. A natural outcome of the construction is a new realization  of deformed Poincar\'e symmetries consistent with Snyder non-commutativity.

In the third section, we construct a field theory on the group $\SO(3,1)^+$ and investigate the different possibilities to constrain the field theory to live on the hyperboloid. We introduce a Fourier transform and a star product to obtain Snyder's non-commutative space. The star product is now associative and the Lorentz sector corresponds to extra  dimensions. We discuss how the Poincar\'e symmetries are deformed in this case.  We show how our construction can be related to the DFR space through a limit \cite{carlson}.

In the fourth section, we discuss some physical aspects of our approaches. In particular we present how the non-associative approach can be obtained from the associative perspective. We also discuss some physical implications to deal with non-associative and non-commutative momenta addition.

In the concluding section, we present some of the many new directions our approach points to.

%%%%%%%%%%%%%%%%%%%%%%%%%%%%%%%%%%%%%%%%%%%%%%%%%%%%%%%%%%%%%%%%%
\section{Curved momentum space} \label{momentum space}
%%%%%%%%%%%%%%%%%%%%%%%%%%%%%%%%%%%%%%%%%%%%%%%%%%%%%%%%%%%%%%%%%

%%%%
\subsection{Momenta and addition}\label{sec:momenta and addition}
%%%%

The Lorentz group $\SO(3,1)$ as a manifold contains two disconnected parts, $\SO(3,1)^+$  and $\SO(3,1)^-$ . The subgroup  $\SO(3,1)^+$  is the set of orthochronous Lorentz transformations  which preserve the time direction.  Using the Cartan decomposition \cite{klymyk}, any element $g\in\SO(3,1)^+$ can be written as a  boost $a$ times a rotation $h$, such that  $g=ah$. We recall that a boost $a$ is parameterized by the boost angle $\eta$ and the  vector $\vec b$, such that $a=e^{i{\eta}\vec b\cdot \vec K} $ where $K_i=J_{0i}$ are the boost generators. The rotation $h$ is given in terms of a angle $\alpha$ and a rotation  vector $\vec r$, such that $h= e^{i{\tilde \eta}\vec r\cdot \vec J}$ in term of the rotation generators $J_i= \demi \epsilon_i^{jk} J_{jk}$.

The coset space $\SO(3,1)^+/\SO(3)$ is defined as the set of equivalence classes $[a]= \lbrace a.h, \,\forall h\in \SO(3)\rbrace$. The boost element $a$ is a representative of the equivalence class $[a]$. We note $L$ the set of boosts $a$, which defines a section for the coset space $\SO(3,1)^+/\SO(3)$.  Both the coset $\SO(3,1)^+/\SO(3)$ and the space $L$ are isomorphic to the upper hyperboloid $\hh^+$.

\medskip

The subgroup $\SO(3,1)^+$ is six dimensional with coordinates given by the angles $(\eta,\alpha)$ and the unit vectors $(\vec b, \vec r)$. Interpreting $\SO(3,1)^+$ as the momentum space, we define the momentum coordinates as functions of these parameters~\footnotemark:
\bes\label{momentum coord coset}
&&\vec k=\ka f_1(\eta)\vec b\nn\\
&&\vec \kkk= \lambda f_2(\alpha)\vec r.
\ees
$\ka$ is the Planck mass ($\hbar=c=1$), associated to the physical 3d momentum, while $\lambda$ is another mass scale, associated to an extra momentum sector and whose physical interpretation is left open at this time.
\footnotetext{Note that instead of introducing $\vec r$, one could introduce the antisymmetric tensor $r^{ij}= \demi \epsilon_k^{ij} r_k$ (and therefore $\kkk^{ij}$) such that
%\be \label{r tensor}
$$
\vec r\cdot \vec J= \demi r^k \epsilon_k^{ij} J_{ij}=  r^{ij} J_{ij}.
$$}
%
%\medskip
%
The group multiplication structure inherited from $\SO(3,1)^+$ induces different types of momenta addition according to the momentum coordinates which we choose.   We use the Cartan decomposition $g_i=a_ih_i$ and write the general product in the group,
\bes\label{sum so31}
 &&a_1 h_1 a_2 h_2= a_1(h_1a_2h_1\mone)h_1h_2= a_{1 2'}\, h_{12'}\, h_1\, h_2= ah\nn\\
 && e^{i{\eta_1}\vec b_1\cdot \vec K} e^{i{\alpha_1}\vec r_1\cdot \vec J}e^{i{\eta_2}\vec b_2\cdot \vec  K}e^{i{\alpha_2}\vec r_2\cdot \vec J}=
 e^{i{\eta}\vec b\cdot \vec  K}e^{i{\vartheta}\vec r\cdot \vec J}
\ees
where we have used the following notations
\bes  \label{loop product 0}
a_1a_2 &=&a_{12}\,h_{12},\quad a_{12}\in L, \, h_{12}\in \SO(3)\\
(h_1a_2h_1\mone)&\equiv& h_1\act a_2= a_{2'}.
\ees
$a_{12}$ is  the new boost obtained from the product of two boosts. We can therefore define a product law on $L$,  as follows:
\be \label{loop product}
a_1\cdot a_2 = a_{12}.
\ee
To determine explicitly the momenta addition induced from \eqref{sum so31}, we need to compute explicitly the terms $a_{12}$ and $h_{12}$, as well as  the sum induced from the group multiplication on $\SO(3,1)$, in terms of the momentum coordinates \eqref{momentum coord coset}. We use the spinorial representation of $\SO(3,1)^+$ in terms of $2\times 2$ group elements in $\SL(2,\C)$ in order to perform the calculations.  4-vectors are represented as $2\times 2$ Hermitian matrices.
\be
(v_0, \vec v_{i})\in\R^4\quad\longrightarrow\quad \left(
\begin{array}{cc}v_0+v_3 & v_1+iv_2 \\ v_1-iv_2 & v_0-v_3\end{array}\right).
%(\alpha_0, \vec \alpha_{i})\in\R^4\quad\longrightarrow\quad \left(
%\begin{array}{cc}\alpha_0+\alpha_3 & \alpha_1+i\alpha_2 \\ \alpha_1-i\alpha_2 & \alpha_0-\alpha_3\end{array}\right).
\ee
Defining the Pauli matrices $\vec{\sigma}$ as
$$
\sigma_1=\left(\begin{array}{cc} 0&1 \\1&0 \end{array}\right),
\quad \sigma_2=\left(\begin{array}{cc} 0&-i \\i &0 \end{array}\right),\quad
\sigma_3=\left(\begin{array}{cc} 1&0 \\0&-1 \end{array}\right),
$$
the spinorial representation is given by $J_i=\f12\sigma_i$ and $K_i=-\f i2\sigma_i$. Thus
group elements in $\SO(3,1)$ are  given by $U=\exp(\vec{u}\cdot\vec{\sigma})$ for arbitrary \textit{complex} vectors $\vec{u}$ and in particular in this representation,  boosts are parameterized as:
$$
a
=e^{i\eta \overrightarrow{b}\cdot\overrightarrow{K}}
= e^{\frac{\eta}{2} \overrightarrow{b}\cdot\overrightarrow{\sigma}}
\,= \cosh\frac{\eta}{2} Id+ \sinh\frac{\eta}{2}
\overrightarrow{b}\cdot \overrightarrow{\sigma}.
$$
Group elements act on $2\times 2$ matrices by conjugation. Defining the origin of the hyperboloid as the 4-vector $v_{(0)}= (1,0,0,0)$, or equivalently as the Hermitian $\one$ matrix, boosts allow to translate it to any point on the hyperboloid:
\be
v\equiv\, a v_{(0)} a^\dagger \,=\, a  a^\dagger
\,=\,(\cosh \eta, \sinh\eta \overrightarrow{b}),\qquad
v_\mu v^\mu=(v^0)^2-(\vec{v})^2=1.
\ee
This definition is clearly invariant under 3d rotations:
$$
ah(ah)^\dagger
\,=\,
ah h^{-1}a^\dagger
\,=\,
a  a^\dagger,
$$
since the spinorial representation provides a unitary representation of $\SU(2)$, i.e $h\dag=h^{-1}$. This gives the explicit isomorphism between the coset space $\SO(3,1)^+/SO(3)$ and the (upper) hyperboloid.
Using this representation, we can compute  the boost $a_{12}$
\be
a_{12}= e^{i {\eta_1}\vec b_1\cdot \vec K} \cdot e^{i{\eta_2}\vec b_2\cdot \vec K}=e^{i {\eta_{12}}(\vec b_1\oplus \vec b_2)\cdot \vec K}.
\ee
Using the definition \eqref{momentum coord coset}, we then deduce a general momenta addition $\vec k_1\oplus \vec k_2=\vec k_{12}$.
\begin{equation}\label{speedcomp}
\overrightarrow{k_{12}}= \overrightarrow{k_1}\oplus\overrightarrow{k_2}= \frac{f(\eta_k)}{\sinh
\eta_k}(C_{k_1}\overrightarrow{k_1}+C_{k_2}\overrightarrow{k_2}), \qquad
\begin{array}{rcl}
C_{k_1}&=& \frac{\sinh\eta_{1}}{f(\eta_{1})}\cosh\eta_{2} +\left( \frac{\sinh\eta_{1}}{f(\eta_1)}\right)^2\frac{\sinh\eta_{2}}{f(\eta_2)}\frac{1}{1+\cosh\eta_{1} }\frac{\overrightarrow{k_1}.\overrightarrow{k_2}}{\ka^2}\\
C_{k_2}&=& \frac{\sinh\eta_{2}}{f(\eta_{2})}.
\end{array}
\end{equation}
We also derive that
\be
\cosh\eta_{12}=\cosh\eta_1\cosh\eta_2+ \frac{\sinh\eta_1}{f(\eta_1)}\frac{\sinh\eta_2}{f(\eta_2)}\frac{\vec k_1\cdot \vec k_2}{\ka^2}.
\ee
We can give the explicit expression of these formulae by choosing specific momentum coordinates. The two canonical choices of coordinates are respectively the \emph{Snyder coordinates} $\vec P$  and the \emph{embedding coordinates} $\vec p$~:
\be
\label{snyder coordinates 0}
\vec P=\ka \tanh \eta \, \vec b \,=\, \ka\,\f{\vec{v}}{v^0},\qquad
\vec p= \ka \sinh \eta\, \vec b \,=\, \ka\,\vec{v}.
\ee
This leads to the following deformed addition of momenta:
\bes
\label{sum snyder}
\vec P_1 \oplus \vec P_2&=& \frac{1}{1+\frac{\vec P_1\cdot \vec P_2}{\ka^2}}\left(\left(1+ \frac{\gamma_1}{1+\gamma_1}\frac{\vec P_1\cdot \vec P_2}{\ka^2}\right)\vec P_1 +  \frac{1}{\gamma_1}\vec P_2\right), \quad\textrm{with}\quad \gamma_1= \frac{1}{\sqrt{1-\frac{\vec P^2_1}{\ka^2}}},\\
(\vec p_1 \oplus \vec p_2)^i&=&(p^0_2+\frac{1}{\ka^2}\frac{1}{1+p^0_1}\vec p_1\cdot \vec p_2)p_1^i +p_2^i, \quad\textrm{with}\quad p^0_d= \sqrt{1+\frac{\vec p^2_d}{\ka^2}}, \quad d=1,2.
\ees
We note $\ominus$ the inverse of this modified sum: $\ominus \vec k\oplus \vec k= \vec k\oplus (\ominus \vec k)= \vec 0$, for $\vec k$ any choice of momentum coordinates. It turns out that this inverse momentum remains trivial:
\bes\label{momentum substraction}
\ominus  \vec k= -\vec k &\Rightarrow& \left(\begin{array}{l}\ominus \vec P = -\vec P \\ \ominus \vec p = - \vec p.\end{array}\right.
\ees
In the product \eqref{loop product 0}, the ``Thomas precession" $h_{12}=e^{{i}\vartheta_{12}\vec r_{12}\cdot \vec J} $ is totally specified by the boosts $a_1$ and $a_2$ and we can compute explicitly:
\be\label{thomas precession}
\tan\frac{\vartheta_{12}}{2} \overrightarrow{r_{12}}
=\frac{\tanh\frac{\eta_1}{2}\tanh\frac{\eta_2}{2}}{1+\tanh\frac{\eta_1}{2}\tanh\frac{\eta_2}{2}
\overrightarrow{b_1}.\overrightarrow{b_2}}(\overrightarrow{b_1}\wedge\overrightarrow{b_2}).
\ee
Using the embedding coordinates $p_i= \ka \sinh \eta\, \vec b$ and the choice $\vec \kkk= \lambda \tan{\f\vartheta 2}\, \vec r$,  we have for example
\be
\kkk_{12}
%\frac{1}{1+\sqrt{1+\frac{\vec \kkk_{12}^2}{\lambda^2}}}\kkk_{12}
= \frac{\lambda}{\ka^2} \frac{\vec p_1\wedge\vec  p_2}{\tilde \gamma_1\tilde \gamma_2 + \frac{\vec p_1\cdot\vec  p_2}{\ka^2}},\quad \tilde \gamma_i= 1+\sqrt{1+ \frac{|\vec p_i|^2}{\ka^2}}.
\ee

Finally, we also compute the product between rotations $h=h_1 h_2$, which  induces a non-trivial addition for the momentum $\kkk$. Still using the spinorial representation $h_i= \cos \f{\alpha_i}{2}\one + i\sin \f{\alpha_i}{2}\, \vec r_i\cdot \vec \sigma$, we have
\bes
\label{rotation comp}
\sin\f{\alpha}{2}\, \vec r&=& \cos\f{\alpha_1}{2}\sin\f{\alpha_2}{2}\, \vec r_2 + \cos\f{\alpha_2}{2}\sin\f{\alpha_1}{2}\, \vec r_1- \sin\f{\alpha_1}{2}\sin\f{\alpha_2}{2} \vec r_1\wedge \vec r_2 \\
\cos\f{\alpha}{2} &=& \cos\f{\alpha_1}{2}\cos\f{\alpha_2}{2} - \sin\f{\alpha_1}{2}\sin\f{\alpha_2}{2}\, \vec r_1\cdot \vec r_2 \nn
\ees
Keeping the same choice of momentum coordinates $\vec \kkk= \lambda \tan{\f\alpha2} \, \vec r$, we get:
%{\bf NEED TO CORRECT THIS FORMULA!!}
\be
\vec \kkk
\,=\,
\f{1}{1-\f1{\lambda^2}\kkk_1\cdot\kkk_2}
\left(\vec \kkk_1 +\vec \kkk_2     - \frac{1}{\lambda} \kkk_1\wedge \kkk_2\right).
%\qquad \Gamma_i= \frac{1}{\sqrt{1+\frac{\vec \kkk_i^2}{\lambda^2}}}.
\ee
Putting together the additions \eqref{speedcomp} and \eqref{rotation comp} allows to determine the general  addition of momenta \eqref{sum so31} inherited from  the group multiplication on $\SO(3,1)^+$.

\subsection{Geometry, measures and Lorentz transformations}\label{lorentz}
We have introduced a time-like vector $v_{(0)}$  in $\R^{4}$, with  little group  $\SO(3)$. By acting with group element $g\in \SO(3,1)^+$ on $v_{(0)}$, we sweep the upper hyperboloid $\hh^+$ which can be seen as embedded in $\R^{4}$
\bes
\hh^+=\{v^\mu\in\R^{4}, \, -v_0^2+v_i^2=-1, \, v_0>0\} \sim \SO(3,1)^+/\SO(3)
\ees
We have chosen to work with the subgroup $\SO(3,1)^+$. We could also extend the construction to the full group $\SO(3,1)$ and consider the coset $\SO(3,1)/\SO(3)$. This case would generate both the upper and lower hyperboloid and we would work with both signs $\pm$ for $v_0$. The natural metric on the (upper) hyperboloid    is Euclidean, so that we are dealing with  the 3d \emph{Euclidean Snyder space-time}. To consider the Lorentzian case, we should consider instead the coset space given either by the de Sitter space $\SO(p,1)/SO(p-1,1)$ or the anti de Sitter space  $\SO(p,2)/SO(p,1)$.

The different  coordinates systems can be defined in terms of coordinates $v_\mu$.  The \emph{Snyder coordinates}  $\vec P$ are simply
\bes \label{snyder coordinates}
&& \ka\frac{v_i}{v_0}= P_i, \quad v_0= \frac{1}{\sqrt{1- \frac{\vec P^2}{\kappa^2}}}
\ees
In this coordinate system, the rest mass is bounded by $\kappa$, which is quite a strong constraint if we try to relate this construction to a physical model. We would like to emphasize however  that this construction is very close to the structure met in Special Relativity \cite{SR}. Indeed in this case, the hyperboloid  is interpreted as the space of 3d velocity $\vec{\rm v} $. The coordinates $v^\mu\in\R^4$ are interpreted as the relativistic speeds. The usual choice of coordinates on the hyperboloid is given by ${\rm v}^i=c\frac{v^i}{v^0}$, which is analogous to the Snyder coordinates. The speed addition is then defined from the product given in \eqref{loop product}  and is non-associative. It is precisely obtained from \eqref{sum snyder} by replacing $\ka$ by $c$ and $\vec P$ by $\vec {\rm v}$.

We also have a simple expression for the \emph{embedding coordinates}  $\vec p$, in which case the rest mass is not bounded:
\bes \label{5d coordinates}
&& \ka v_i=  p_i, \quad v_0=  \sqrt{ \frac{\vec p^2}{\ka^2} +1}.
\ees

%\medskip

Before discussing the construction of a field theory on the hyperboloid interpreted as the momentum space, we would like to discuss the notion of  symmetries.
Comparing our 3d Euclidean Snyder space-time toy model to an actual 4d Lorentzian Snyder space-time, our $\SO(3)$ transformations will play the role of Lorentz transformations, while our boosts in $\SO(3,1)^+/\SO(3)$ will play the role of (deformed) Poincar\'e translation operators.
%By analogy to the Lorentzian case, we shall call the  $\SO(3)$ transformations  the "Lorentz transformations", even if  we are dealing with the Euclidian transformations (\ie the rotations)  in our example.
%
The subgroup $\SO(3)$ acts naturally by the adjoint action on itself. Using the Cartan decomposition, we have that
\be
g\,\dr \, h_1 \,g \, h_1\mone= (h_1 \,a h_1\mone) \, (h_1 \, h  \, h_1\mone).
\ee
We consider therefore the natural adjoint action of $\SO(3)$ on  itself and on $\SO(3,1)^+/\SO(3)$. In particular the rotations acts simply on $\vec r$:
\be\label{lorentz on r}
[J_{ij},r_l]=\delta_{jl}r_i-\delta_{il}r_j.
\ee
Moreover, the symmetry transformations of $\SO(3,1)^+/\SO(3)$ can be also deduced from the induced transformations of $\R^{4}$ in which the hyperboloid is embedded. The coordinates $v_\mu$  transform linearly under $\SO(3)$, and it is easy to see that the Snyder coordinates will then in turn also transform linearly, under the adjoint action. For example for the hyperboloid, we have
\begin{eqnarray}\label{poincare alegbra snyder}
&& [J_{ij},p_l]=\delta_{jl}p_i-\delta_{il}p_j, \nn\\
&& [J_{ij},P_l]=\delta_{jl}P_i-\delta_{il}P_j.
\end{eqnarray}
The invariant Haar measure $[dg]$ on the group $\SO(3,1)^+$ is of course invariant under the adjoint action of $\SO(3)$ and so is the invariant Haar measure $[dh]$ on $\SO(3)$. Since the we are using the Cartan decomposition, the measure $[dg]$ can be split into the product of the measure on $\SO(3,1)^+/\SO(3)$ and the measure on the subgroup $\SO(3)$:
\be
[dg]= [da][dh].
\ee
The measure $[da]$  can be expressed in terms of the embedding coordinates:
$$[da] = \ka^{3}\,d^{4}v\, \theta(v_0)\,\delta (v^\mu v_\mu +1),$$
where $\theta(v_0)$ imposes the condition $v_0>0$.
From this expression it is obvious that the measure $[da]$  is  invariant under the adjoint action of $\SO(3)$. Writing the measure in terms of the Snyder coordinates \eqref{snyder coordinates}  or the embedding coordinates \eqref{5d coordinates} on the 3d hyperboloid, we have
\bes\label{measure hyperboloid}
[da]\,=\,
\ka^3\,d^4v \,\theta(v_0)\,\delta(v_0^2-v_i^2-1)
\,=\,[dk]\equiv
\left(\begin{array}{l}
[dP]= d^3P\, \left({1- \frac{\vec P^2}{\kappa^2} }\right)^{-2}\\
{[}dp{]}= {d^3p}\left({{1+ \frac{\vec p^2}{\kappa^2}}}\right)^{-\demi},
\end{array}\right.
\ees
The measure  on $\SO(3)$ is the usual Haar measure, which is the measure on the 3-sphere (quotiented by $\Z_2$) inherited from the Lebesgue measure on $\R^4$~:
$$[dh] = \lambda^{3}\,d^{4}w\,\theta(w_0)\, \delta (\sum_{A=1,..,4}w_A^2 -1), \qquad w_A=(\cos\f\alpha2,\sin\f\alpha2\,\vec{r})\in \R^4, $$
and expressed for example in terms of the momentum variable $\kkk= \tan\f\alpha2 \, \vec r$, it becomes
\be\label{measure sphere}
[dh]= \frac{d^3\kkk}{\left(1+\frac{\vec \kkk^2}{\lambda^2}\right)^2}.
\ee

\section{Scalar field theory on the hyperboloid  as momentum space}
We are now interested in constructing a scalar field action defined on momentum space which is the hyperboloid $\hh$.

We note $k$ the choice of coordinates on the hyperboloid, which can be either the Snyder coordinates $P$ in \eqref{snyder coordinates} or the embedding coordinates $p$ in \eqref{5d coordinates}.

\subsection{Action in momentum space: non-associative convolution product}\label{sec:field on coset}
The scalar field $\phi$ is a function $\phi(a)\in C(\hh)$ seen as a distribution. Another useful distribution is the Dirac function $\delta(a)$ on the hyperboloid $\hh^+$ . This $\delta$-distribution can be constructed from the delta function on  the group
$$
\delta_a(b)\equiv\delta_{\hh^+}(b\cdot a\mone)=\int [dh]\,\delta_{\SO(3,1)^+}(ba\mone h), \,b\in L. %=\delta^{(3)}(\vec{k}),
$$
where we have also used the  product \eqref{loop product}.
In the following, we will drop the indices $\hh$ or $\SO(3,1)$.
We can always choose a momentum coordinate system $k_\mu$ and express the field  in this coordinate system $\phi(a)\sim \phi(k_i)$. Since the measure factor \eqref{measure hyperboloid} is trivial at the origin $k=0$, the Dirac distribution is also simply
$\delta(a)\sim \delta^{3}(k_i)$.

The convolution product is the natural product between distributions.
We note this product $\dia$ and introduce it in a similar way as the  convolution product when dealing with a group, except that now we use the  product $a_1\cdot a_2=a_{12}$.
\bes \label{convolution}
&& \phi\dia \phi(a)\equiv \int [da_1][da_2]\, \phi(a_1)\phi(a_2)\int [dh]\, \delta(a\mone (a_1a_2) h )= \int [da_1][da_2]\, \phi(a_1)\phi(a_2)\, \delta(a\mone\cdot (a_1\cdot a_2) ),\\
&& \delta_{a_1}\dia \delta_{a_2}\equiv \delta_{a_1\cdot a_2}= \delta_{a_{12}}.
\ees
We emphasize   that the $\delta$-distribution in the left hand side of the first line, is on the $\SO(3,1)$-group, while the $\delta$-distribution on the right hand side is on the coset.
In the chosen coordinates system given by $k$, the Dirac distribution $\delta(a_{12})$ can be rewritten as $\delta(k_1\oplus k_2)$. Since we are using the  product \eqref{loop product}, the convolution product will also be non-associative and we need to specify the order in which we perform the products.
\bes
(\phi\dia(\phi \dia \phi))(a)&\equiv& \int [da_i]^3\, \phi(a_1)\phi(a_2)\phi(a_3)\int [dh_i]^2\, \delta(a\mone( a_1(a_2a_3)) h_1h_2 )\nn \\
&=& \int [da_i]^3\, \phi(a_1)\phi(a_2)\phi(a_3)\delta(a\mone\cdot a_{1(23)}), \qquad a_{1(23)}= a_1\cdot(a_2\cdot a_3)
\ees

\medskip

We can now construct an action for a field defined on $\SO(3,1)^+/\SO(3)$, interpreted as momentum space. We consider the  measure $[da]$, and some propagator  $\kk(a)$ and % that is invariant under the action of $H$.
 define a $\phi^3$ type scalar field action as
\bes\label{action field}
\ss(\phi)&=& \int [da]^2\, \phi(a_1)\kk(a_1)\phi(a_2)\delta(a_1\cdot a_2)+ \frac{\lambda}{3!} \phi\dia(\phi\dia\phi) (e)\nn\\
&=& \int [dk]^2\, \phi(k_1)\kk(k_1)\phi(k_2)\,\delta(k_1\oplus k_2)+ \frac{\lambda}{3!} \int [dk]^3\,\phi(k_1)\phi (k_2)\phi(k_3)\,\delta(k_1\oplus(k_2\oplus k_3)).
\ees
When using the coordinates $k$, this action resembles very much the usual scalar field  action. The  differences come in the measure which encodes now the fact that we are working on the hyperboloid and in the conservation of momenta which uses the modified (non-associative) addition of momenta.\\

We can check that our action is invariant under the Lorentz transformations given by the group $\SO(3)$.  We have seen in  subsection \ref{lorentz} that $\SO(3)$ acts by the adjoint action on the coset $a\dr h\act a= h\mone a h$, $h\in \SO(3)$.
On the other hand, the scalar field and the Dirac delta function transform by definition as scalars, so we have
\bes \label{lorentz snyder}
\phi(a)&\dr& \phi (h\act a) \\
\delta(a)&\dr& \delta (h\act a).
\ees
The convolution product between functions (resp. Dirac functions) gives a function (resp. a Dirac function). It transforms under $\SO(3)$   as a (scalar) function (resp. a Dirac function).  We have for example
\bes\label{lorentz on convolution snyder}
 h\act(\phi \dia \psi ( a))&=&\phi \dia \psi ( h\act a) \\
h\act \delta(a_1\cdot a_2)&=&  \delta(h\act (a_1\cdot a_2))=\delta( (h\act a_1)\cdot (h\act a_2)).
\ees
In particular, one notes that the convolution product evaluated at the identity $e$ of the coset, which corresponds to zero momentum $k_\mu=0$, is therefore invariant under the Lorentz action. Considering a $\phi^3$ interaction term for example, we have
\be\label{conv product inv}
\phi\dia(\phi\dia \phi) (e) \dr\phi\dia(\phi\dia \phi) (h\act e)=\phi\dia(\phi\dia \phi) (e)
\ee
Finally, if the propagator $\kk(a)$ is invariant under the Lorentz transformations $\kk(h\act a)= \kk(a)$, it is clear that the action \eqref{action field} is invariant under the adjoint action of $\SO(3)$.

\medskip

Before ending the section we would like to comment on the choice of momentum. We have freedom in the choice of momentum. For example we can take the Snyder choice $P_i$,  given by the Snyder coordinates \eqref{snyder coordinates}. In this case the propagator can be taken to be
\be \label{snyder propa 1}
\kk(a)= P^2-m^2= \ka^2 \frac{v^iv_i}{v_0^2} -m^2.
\ee
This is properly invariant under $\SO(3)$ rotations since $\SO(3)$ acts linearly on $P_\mu$.
On the other hand we can also consider  the embedding coordinates $p_i$, with a propagator
\be \label{snyder propa 2}
\kk(a)= p^2-m^2= \ka^2 {v^iv_i} -m^2.
\ee
This is also invariant under $\SO(3)$ rotations, although it differs importantly from the earlier choice.

Contrary to the $\ka$-Minkowski case \cite{sitarz} or the $su(2)$ space \cite{battista}, we are not aware of any study of the non-commutative differential calculus in the Snyder space-time. It is indeed the choice of differential calculus that tells us which kind of momentum which should choose (see for example \cite{amelino} and \cite{laurent-kowalski}).  This calculus deserving some specific attention on its own, we leave it on the side for now.

\subsection{Action in Snyder space-time: non-associative star product } \label{sec:action non assoc}
We construct now a generalized Fourier to define the dual space-time. We introduce first the c-numbers $x_i\in \R^3$. Given a choice of momentum coordinates $k$, we consider   the plane-wave  $e^{ik\cdot x}=e^{ik(a)\cdot x}$ and the star  product noted $\divideontimes$ between the plane-waves which  introduces the modified momenta additions \eqref{speedcomp}.
\bes \label{product planewave}
e^{ik_1\cdot x }\st e^{ik_2 \cdot x}\equiv e^{i(k_1\oplus k_2)\cdot x }&\dr&  \left(\begin{array}{l}e^{iP_1\cdot x }\star e^{iP_2\cdot x }\equiv e^{i(P_1\oplus P_2)\cdot x }\\
e^{ip_1\cdot x }* e^{ip_2\cdot x }\equiv e^{i(p_1\oplus p_2)\cdot x }.\end{array}\right.
\ees
We define the Fourier transform of a distribution $\phi$ as
\be
\hat \phi(x)\equiv
\int [da] \,  e^{ik(a)\cdot x  } \phi(a)
\,=\,
\int [dk] \,  e^{ik\cdot x  } \phi(k),
\ee
in particular we define a deformed ``fuzzy" $\delta$-distribution and we can write the space-time coordinate as derivation operators in the momentum space:
\be
\delta_\st(x)= \int [dk]\, e^{ik\cdot x }, \qquad
x_i= -i\int [dk]\, \delta(k)\,\partial_{k^i} e^{ik\cdot x}. \label{x snyder}
\ee
This fuzzy $\delta$-function should truly be considered as a distribution and is a priori not equal to the standard $\delta$-distribution on $\R^3$ due to the different measure $[dk]\ne d^3\vec{k}$.
%
%As an example, choosing the Snyder momentum coordinate, we will have:
%\be
%\delta_\st(x)
%\,= \,
%\int_{P^2\le \ka^2} \f{d^3P}{\left(1-\f{P^2}{\ka^2}\right)^2}\, e^{i\vec{P}\cdot \vec{x} }
%\,= \,
%\f{1}{|x|}\int_{P^2\le \ka^2} \f{4\pi P dP}{\left(1-\f{P^2}{\ka^2}\right)^2}\, \sin P|x|
%\,= \,
%\f{1}{|x|}\int_0^{\f\pi2} \f{4\pi \sin\theta d\theta}{\cos\theta}\, \sin P\sin\theta,
%\ee
%%This expression is actually divergent for $\theta\arr\f\pi2$.
%%
%For the embedding coordinates, we have:
%\be
%\delta_\st(x)
%\,= \,
%\int \f{d^3p}{\sqrt{1+\f{p^2}{\ka^2}}}\, e^{i\vec{p}\cdot \vec{x} }
%\,= \,
%\f{1}{|x|}\int_0^\infty \f{4\pi p dp}{\sqrt{1+\f{p^2}{\ka^2}}}\, \sin p|x|.
%\ee

%The inverse Fourier transform is defined in turn as
%\be
%\phi(k)= \int[dx]\,  e^{i(\ominus k)\cdot x } \st \hphi(x)= \int[dx]\,  e^{-i k \cdot x } \st \hphi(x),
%\ee
%in particular we have
%\bes
%&& \delta (k)= \int d^3x \,   e^{-ik\cdot x}, \qquad
%k^\mu = i\int  d^3x \, \delta(x) \st \partial_{x_{\mu}} e^{-i k\cdot x}. \label{def p dsr}
%\ees

As usual, the $\st$ product can be seen as the dual of the convolution product.
\bes
\int [da]\,  e^{ik(a)\cdot x}  (\psi\dia \phi)(a) &=&\int [dk_1][dk_2]\, \psi(a_1)\phi(a_2)\int [dk][dh]\,   e^{ik(a)\cdot x}  \delta(a\mone a_1a_2 h )\nn\\
&=&  \int [dk_1][dk_2]\, \psi(k_1)\phi(k_2)   e^{i(k_1\oplus k_2)\cdot x  } \nn \\
&=& \int [dk_1][dk_2]\, \psi(k_1)\phi(k_2)   e^{i k_1\cdot x  }\st e^{ik_2 \cdot x } \nn\\
&=&  \hat \psi \st \hat \phi (x)
\ees
Since the convolution product is non-associative, the $\st$ product will also be non-associative, so that one needs to be careful with the order in which we group the terms. A direct Fourier transform shows that one has for example
\be
\phi\dia(\phi\dia\phi)(e) \dr \int [d^3x]\, \phi\st(\phi\st\phi)(x).\ee

To get a better grasp of the $\st$ product,  we  first study the product between monomials of first order $x_{i}\st x_{j}$.
\bes\label{1st order monomial}
x_{i} \st x_{j}&=& \left(\begin{array}{l}x_{i} \star x_{j}= - \int[dP]^2\, \delta(\vec P_1)\,  \delta(\vec P_2)\,\partial_{P_1^{i}}\partial_{P_2^{j}} e^{i(P_1\oplus P_2)\cdot x} =x_{i}\,x_{j} \\
\\
x_{i} * x_{j}=- \int[dp]^2\, \delta(\vec p_1)\,  \delta(\vec p_2)\,\partial_{p_1^{i}}\partial_{p_2^{j}} e^{i(p_1\oplus p_2)\cdot x}=x_{i}\,x_{j}.\end{array}\right.
\ees
We have therefore the perhaps surprising result that the coordinate functions commute with each other:
\be\label{comutative snyder}
\com{x_{i},x_{j}}_\st= 0.
\ee
However, due the lack of associativity, this apparent commutativity does not extend to more complicated functions other coordinates. In particular, we do not expect that the product of a monomial of degree one with a monomial of degree two is the same as the product of the monomial of degree two with the monomial of degree one. For instance, we compute:
%  The triple product of the type $x_{i}\st(x_{j} \st x_{m})$ is less trivial
\bes 
x_{i}\st (x_{j} \st x_{m})&=& i\int[dk]^3\,\delta(\vec k_1)\, \delta(\vec k_2)\,\delta(\vec k_3)\, \partial_{k_1^{i}}  \partial_{k_2^{j}}\partial_{k_3^{m}}e^{ix\cdot(k_1\oplus(k_2\oplus k_3))}\nn\\
&=& \left(\begin{array}{l} x_{i} \star (x_{j}\star x_{m})= -\frac{1}{\ka^2}\delta_{{j}{m}}x_{i} +x_{i} x_{j} x_{m}\\
\\
x_{i} * (x_{j}*x_{m})= \frac{1}{\ka^2}\delta_{{i}{m}}x_{j} +x_{i} x_{j} x_{m}\end{array}\right.
\label{triple product}
\ees
The difference between the two choices of star product is manifest in the contribution of order $\ka^{-2}$. We can explicitly check that the star product is non-associative. We have for example
$$
(x_{i} * x_{j})* x_{m} = -\frac{1}{\ka^2}x_{m}\delta_{{i}{j}}+x_{i} x_{j} x_{m}\neq x_{i} * (x_{j}*x_{m}).$$
The $\st$ product encodes therefore some type of non-commutativity  in a slightly different way than usual. We still want to identify the non-commutative space-time behind this $\st$ product. For this, we introduce the   position operator which  acts by ($\st$-)multiplication by the coordinate $x_{i}$.
\be \label{def X snyder}
X_i \act \hat f(x)= x_i \st \hat f(x).
\ee
Using the Fourier transform and either of the coordinates $\vec P$ or $\vec p$, we can explicitly calculate the action of the commutator  $[X_i,X_j]$ on a function $\hat f(x)$, using either of the coordinates $\vec P$ or $\vec p$:
\be
[X_i, X_j] \act\hat f(x)= \left(X_i X_j- X_j X_i\right) \hat f(x)= x_i\st(x_j \st \hat f(x))- x_j\st(x_i \st \hat f(x))=i\frac{1}{\ka^2} J_{ij}\act \hat f(x).
\ee
This result can be checked explicitly when considering $f(x)=x_l$, for a given  $l$ and the results given in \eqref{triple product}.
This shows that the commutator of the  position operators does satisfy
\be \label{snyder bracket}
[X_i,X_j]=  i\frac{1}{\ka^2}J_{ij},
\ee
which encodes the commutation relation of the Snyder coordinates \cite{snyder}. It is important to underline the difference between this key commutation relation and the apparent commutativity of the coordinates \eqref{comutative snyder}.
\emph{We have therefore constructed  a realization of the (Euclidian) Snyder space-time in terms of a star product}. The position operators $X_{i}$ are realized as
\be \label{X snyder boost}
X_i=\frac{1}{\ka} J_{0i},
\ee
and we can  check explicitly by introducing the action of the $J_{0i}=K_i$ on the functions on momentum space and using the Fourier transform that the operator $K_i$ does indeed act by $\st$-multiplication by $x_{i}$ as in \eqref{def X snyder}.

\medskip

We have now the tools to determine the action \eqref{action field} as an action in expressed in a non-commutative space-time.
Using  that
\be
-i \partial_{i} \,  e^{ik\cdot x} = k_{i} e^{ik\cdot x},\ee
and making the Fourier transform of the action \eqref{action field} gives
\bes \label{action non associative position}
\ss(\phi)&=& \int [dx]\,\left( (\partial_{i}\hphi\st \partial_{i}  \hat\phi)(x)+ m^2 (\hat \phi\st \hat \phi)(x)+\frac{\lambda}{3!}( \phi\st(\hat \phi\st \hat \phi))(x)\right)
\ees
This action for a scalar field looks very similar to the scalar field action introduced in non-commutative space-times such as Moyal and $\ka$-Minkowski. The main difference is that one needs to be careful in the way we group the product of fields, due to the non-associative property of the $\st$ product.

\subsection{Deformed   Poincar\'e symmetries}
We introduce now the symmetries of this non-commutative space-time. The Lorentz transformations can be determined by duality using the Fourier transform. We then introduce the translations compatible with the non-commutative structure.

The action of the Lorentz group on functions on space-time can be determined by Fourier duality.
\be
h\act \hphi(x)\equiv \int [dk]\, \phi(h\act k) e^{ik\cdot x}
\ee
We consider an infinitesimal transformation $h\sim 1+i\epsilon\cdot J$, and make use of the  invariance of the  measure $[dk]$ under the adjoint action of $H$:
\be
(1+\epsilon\cdot J)\act \hphi(x)\equiv \int [dk]\, \phi(h\act k) e^{ik\cdot x}
= \int [dk]\, \phi(k) e^{i(h\mone\act k)\cdot x} \sim \int [dk]\, \phi(k)(1+ x \cdot ((\epsilon \cdot J) \act k))e^{ix\cdot k}.
\ee
This allows in particular to define the (infinitesimal) Lorentz transformations action on the coordinates $x_{i}$.
\bes \label{rotation on x}
J_{ij}\act x_l &\equiv& - i \int [dk] \,  \delta(J_{ij}\act k_l) \partial_{(J_{ij}\act k_l)}  e^{i k\cdot x}\nn\\
&=&i(\delta_{jl}x_i- \delta_{il}x_j),
\ees
where we used the transformations \eqref{poincare alegbra snyder}.
As one could have guessed $\SO(3)$ is acting in the standard way in the space-time coordinates, and we check easily that the action \eqref{action non associative position} is indeed invariant under the Lorentz transformations expressed in space-time.

Now that we have defined space-time, we can construct the notion of translations and check that the action  is indeed also invariant under the translations. To this aim we define how  the translations act on the field expressed in momentum space.
 \bes
\hphi(x+\epsilon)&=& \int [da]\, \phi(a)  e^{ik(a)\cdot(x+\epsilon) }\nn\\
&\Rightarrow& \phi(a)\,\dr\, e^{ik(a)\cdot \epsilon} \phi(a)\\
(\hphi \st \hphi)(x+\epsilon)&=& \int [da_i]^2\, e^{ik(a_1\cdot a_2)\cdot (x+\epsilon) }\, \phi(a_1)\phi(a_2)  \nn\\
&\Rightarrow& \phi(a_1)\phi(a_2)\, \dr \, e^{ik(a_{1})\cdot\epsilon  }\st e^{ik(a_{2})\cdot\epsilon  }\, \phi(a_1)\phi(a_2) .
\ees
We see therefore that in momentum space the translations act by  a  phase multiplication.
%which is consistent with the transformations \eqref{translation snyder}.
Moreover, when dealing with many  fields, we  use the $\st$ product between the plane-waves to define  the transformation of the product of field. Thanks to the Dirac delta function encoding the conservation of momentum, it is immediate to check that the action is invariant under  translations as well. We have therefore identified a deformation of the Poincar\'e symmetries which is consistent with Snyder's commutations relations. This new deformation has not appeared when dealing with the classification of the Poincar\'e symmetries deformations in \cite{zak}, since these deformations were still encoding  a (co-)associative feature.

\section{Field theory on  $\SO(3,1)^+$ as momentum space }
Instead of using the product \eqref{loop product}  on the hyperboloid, we are going to use  the full group structure and choose an adequate scalar field and/or modify the group convolution product such that the final field theory is defined  on the space $\SO(3,1)^+/\SO(3)$. We shall then construct the Fourier transform from  $C(\SO(3,1)^+)$ to $C_*(\R^6)$ to identify how the chosen field theory can actually be interpreted as a field theory on a non-commutative space-time of the Snyder type. We shall discuss then how this construction can be related to the non-commutative Doplicher-Fredenhagen-Roberts space.

\subsection{Action in momentum space: group convolution product} \label{sec:field on group}
 We consider  the Dirac  delta function on \emph{the group} $\SO(3,1)^+$
$$\delta(g)\sim \delta(a)\delta(h)\sim \delta^3(k_i)\delta^3(\kkk_j),$$ and the scalar field $\psi(g)=\psi(a,h)$ defined on $\SO(3,1)^+$.  We have the convolution product on the group given by the standard formula
\bes\label{G convolution 2}
&&f\circ \tilde f(g)= \int [dg]^2\, f(g_1)\tilde f(g_2)\delta(g\mone g_1g_2)\\
&& \delta_{g_1}\circ \delta_{g_2}= \delta_{g_1g_2}.
\ees
We can construct an scalar field action in a straightforward manner, generalizing the construction on the groups $\SU(2)$ \cite{noui} and $\AN_p$ \cite{laurent-kowalski}.
 We introduce a  propagator  $\kk(g)$ and   define a $\psi^3$ type scalar field action as
\bes\label{action field assoc}
\ss(\psi)&=& \int [dg]^2\, \psi(g_1)\kk(g_1)\psi(g_2)\delta(g_1g_2)+ \frac{\lambda}{3!} \int [dg]^3\, \psi(g_1)\psi(g_2)\psi(g_3)\delta(g_1g_2g_3)
%\\
%&=& \int [dk]^2\, \phi(k_1)\kk(k_1)\phi(k_2)\,\delta(k_1\oplus k_2)\delta(\kkk_{12})+ \frac{\lambda}{3!} \int [dk]^3\,\phi(k_1)\phi (k_2)\phi(k_3)\,\delta((k_1\oplus k_2)\oplus k'_3)\delta(\kkk_{123'}\oplus \kkk_{12}).\nn
\ees
%We emphasize that contrary to the  previous section, we are using  an associative product.
Following the construction of scalar field theory on $\SU(2)$ \cite{noui}, it is quite clear that this scalar field action will be invariant under the Drinfeld double $\dd(\SO(3,1)^+)$. We are going to show in the following how $\SO(3,1)^+$ acts as a symmetry and in the next section how $\SO(3,1)^+$  encodes the deformed translation symmetry. $\dd(\SO(3,1)^+)$ is  not a deformation of the Poincar\'e group acting on $\R^6$, since the Lorentz group contribution should be $SO(5,1)$. For completeness, one should show either that the Drinfeld double of $\SO(3,1)^+$  is the largest symmetry for this action or that it exists a deformation of the Poincar\'e group acting on $\R^6$, which could be identified as the fundamental symmetry of the action. We postpone these investigations and show how $\SO(3,1)^+$ is a symmetry of this action, which will tell us in particular how the  group $\SO(3)$ acts.

There is a natural action of $\SO(3,1)^+$ on itself, given by the adjoint action. The field $\psi$ transforms then as
\be
\psi(g)\dr \psi(\tilde{g}g\tilde g \mone).
\ee
The convolution product of functions, evaluated at the identity is therefore clearly invariant under such transformation. Moreover, provided that the propagator is invariant under the adjoint action of $\SO(3,1)^+$, it is clear that the action\eqref{action field assoc} is invariant under the adjoint action of $\SO(3,1)^+$. In particular the action is therefore invariant under the adjoint action of the subgroup $\SO(3)$. If the propagator is only invariant under $\SO(3)$ , $\kk(g)= \kk(hgh\mone)$ $\forall h\in \SO(3)$,  the Lorentz symmetry of the action is then broken to $\SO(3)$.

Finally let us remark that we have again some freedom in choosing the relevant momentum variables (as discussed at the end of section \ref{sec:field on coset}). We are not aware of any explicit classification of the differential calculus for the non-commutative space given by $\sll(2,\C)$, so that we leave this issue for later investigations.

\medskip

We want to identify different types of field and modification of this action such that we recover an action defined on the hyperboloid seen as a subspace of $\SO(3,1)^+$:
\begin{itemize}
\item{Field restriction:} \begin{itemize}
\item We can restrict our field to live on the coset i.e impose a $\SO(3)$-invariance of the field, $\psi(g)=\psi(gh), \forall h\in\SO(3)$. Then the field is defined by its section on the hyperboloid, $\psi(ah)\equiv\phi_1(a)$. Assuming an invariance in momentum space means the field is strongly localized in the dual space-time coordinates.
\item  The alternative is to localize the field on the hyperboloid as a distribution $\psi(g)\equiv\phi_2(a)\delta(h)$. Assuming a strong localization in momentum space amounts to a invariance along the dual space-time coordinates.
\end{itemize}
\item{Action modification:} We can insert a damping factor $f(g)$ in the action which would damp or even localize the action on the relevant sector. At the same time we can also  use the previous field restrictions.
\end{itemize}

 Let us first consider the action \eqref{action field assoc} with the choice $\psi(g)=\phi_1(a)$ and a choice of propagator  that does not depend on the $\SO(3)$ sector, \ie $\kk(g)=\kk(a)$.
\bes\label{action field assoc 1}
\ss_1(\phi_1)&=& \int [da]^2[dh]^2\, \phi_1(a_1)\kk(a_1)\phi_1(a_2)\delta(a_1h_1a_2h_2)+ \frac{\lambda}{3!} \int [da]^3[dh]^3\, \phi_1(a_1)\phi_1(a_2)\phi_1(a_3)\delta(a_1h_1a_2h_2a_3h_3)
\ees
The choice of field restriction and propagator is clearing breaking the Lorentz symmetry of the action down to $\SO(3)$, which is acting by adjoint action.

 To understand the structure of this scalar field action,  let us focus for simplicity  on the mass term,  expressed in a coordinate momentum $k$.
\be
\int [dk]^2\,  \phi_1(k_1)\phi_1(k_2)\int [dh]\,\delta( k_1\oplus h\rhd k_2),
\ee
where the sum $\oplus$ is the modified sum inherited from \eqref{speedcomp}.
First, we see that the integral over $h$ averages over the direction of the second momentum $\vec{k}_2$, thus the convolution loses all information about this direction and only remembers the modulus of $\vec{k}_2$. In other words, the mass term does not depend on the entire field $\phi_1(k)$ but only on its radial component $\tl{\phi_1}(k)=\int dh \phi_1(h\rhd k)$. Second, the momentum conservation can not be expressed simply in term of a deformed addition of momenta. Actually conservation of the momentum $k$ is definitely violated. At the end of the day, we obtain a non-trivial distribution of the resulting (final) momentum $\vec k$ which not only depends on the initial momenta $k_1$ and $k_2$ but also on the arbitrary group rotation $h$. As a final comment, we emphasize that the momentum conservation induced here is given in terms of an associative sum, even though using the non-associative sum \eqref{speedcomp}. The presence of the $h$ re-establishes  the associativity.

\medskip

The second possibility is to consider $\psi(g)\equiv\phi_2(ah)\delta(h)$. In this case, the action \eqref{action field assoc} becomes
\bes\label{action field assoc 2}
\ss(\phi_2)&=& \int [da]^2\, \phi_2(a_1)\kk(a_1)\phi_2(a_2)\delta(a_1\cdot a_2)\delta(h_{12})\nn \\ &&+ \frac{\lambda}{3!} \int [da]^3\, \phi_2(a_1)\phi_2(a_2)\phi_2(a_3)\delta(a_1\cdot(a_2\cdot a_3))\delta(h_{1(23)}h_{12}).
\ees
Once again it is straightforward to notice that our choice of field and propagator  breaks the Lorentz covariance to $\SO(3)$.  The terms $h_{12}$ and $h_{1(23)}$ are the Thomas precession contributions obtained from the multiplication of boosts as in \eqref{loop product}. These terms are totally determined in terms of the $a_i$. In fact it is not complicated to see that  the Dirac delta function on the $a_i$ and the one on the $h$ are not independent, so that integrating one $a_i$ leads to a $\delta(0)$ for the $h$ sector. In case of a compact group, as we are considering here, $\delta(0)$ is not divergent \cite{noui, etera}. However, when dealing with a non-compact group, $\delta(0)$ is divergent as usual and this choice $\psi(g)\equiv\phi_2(ah)\delta(h)$   leads to divergences and a badly defined theory. Maybe this could be cured by using half-density distribution instead of the straightforward $\delta$-distribution. But we do not investigate further in this direction.

\medskip

We now consider the  alternative consisting in introducing a damping term in the action. We consider the damping function $f(g)$ and the action defined first  for the field $\psi$ on $\SO(3,1)^+$.
\bes\label{action field assoc 3}
\ss^f(\psi)&=& \int [dg]^3\, \psi(g_1)\kk(g_1)\psi(g_2) f(g_3)\delta(g_1g_2g_3)+ \frac{\lambda}{3!} \int [dg]^4\, \psi(g_1)\psi(g_2)\psi(g_3)f(g_4)\delta(g_1g_2g_3g_4)
\ees
Provided that $f(g)$ is invariant under the  transformations $\SO(3,1)^+$, this group will still be a symmetry of the action.
A typical choice for $f(g)$ would be $f(g)= \delta(a)$, with $g=ah$. This choice breaks the Lorentz symmetry of the action down to $\SO(3)$. The conservation of momenta in \eqref{action field assoc 3} is similar to the one appearing in the action \eqref{action field assoc 1}, with the presence of an extra integration over the group $\SO(3)$.
\bes\label{action field assoc 31}
\ss^f(\psi)&=& \int [da]^2[dh]^3\, \psi(a_1,h_1)\kk(a_1,h_1)\psi(a_2,h_2) \delta(a_1h_1a_2h_2h_3)\nn\\
&& + \frac{\lambda}{3!} \int [da]^3[dh]^4\, \psi(a_1,h_1)\psi(a_2,h_2) \psi(a_3,h_3)delta(a_1h_1a_2h_2a_3h_3h_4).
\ees
In our example, since the group is compact, the extra integration will not affect the construction and the momenta conservation are essentially the same as in \eqref{action field assoc 1}. In fact introducing also the field restriction $\psi(g)\equiv \phi_1(a)$ and a propagator $\kk(g)=\kk(a)$, one recovers exactly \eqref{action field assoc 1}. If dealing with a non-compact group, we have divergences that plague the theory. We can also consider the field restriction $\psi(g)\equiv \phi(a)\delta(h)$, in which case, we recover the non-associative field theory \eqref{action field}.

\subsection{Action in  Snyder space-time: associative star product}
We introduce   the coordinates $(x_i,y_j)\in \R^{6}$.  An alternative choice of variable for the $y$ sector is given by the antisymmetric tensor
 $j_{ij}$, such that $y_j=\demi \epsilon_j^{mn}j_{mn}$. Given an element $g\in \SO(3,1)^+$, we introduce the plane-wave $e_g(x,y)$. The product of plane-waves is encoded as
 \be
(e_{g_1}\circledast e_{g_2})(x,y)\equiv  e_{g_1g_2}(x,y).
\ee
We recall that we are considering the Cartan decomposition $g=ah$. This decomposition singles out some specific cases of interest for the product.
\bes
(e_{h_1}\circledast e_a)(x,y)&=& (e_{h_1ah_1\mone}\circledast e_{h_1})(x,y), \quad h_1\in \SO(3), \, a\in \SO(3,1)^+/\SO(3)\label{noncom planewave}\\
(e_{a_1}\circledast e_{a_2})(x,y)&=& e_{(a_1\cdot a_2) h_{12}}(x,y) \quad a_i\in \SO(3,1)^+/\SO(3), \, h_{12}\in \SO(3). \label{prod moyal snyder}
\ees
The last case resembles  the Moyal case, as it can be made explicit by  introducing a  choice of momentum  $(k,\kkk)$ given from \eqref{momentum coord coset}. The plane-wave is then defined as
 \be
e_{k,\kkk}(x,y)\equiv e^{i (k\cdot x+ \mathfrak{k}\cdot y)}= e^{i( k\cdot x+\mathfrak{k}^{mn} j_{mn})}.
\ee
The  product of plane-wave  is inferred from the modified sum \eqref{sum so31}
\bes \label{prod of planewave}
(e_{k_1,\kkk_1} \circledast e_{k_2,\kkk_2})(x,y)= e_{k_1\oplus k'_2,\kkk_{12'}\oplus\kkk_1\oplus \kkk_2}(x,y), \qquad k'_2= h_1\act k_2.
\ees
The product in \eqref{prod moyal snyder} becomes then
\be \label{simple product}
e^{i k_1\cdot x} \circledast e^{i k_2\cdot x}= e^{i (k_1\oplus k_2)\cdot x} e^{i \kkk_{12}\cdot y},
\ee
where $\kkk_{12}$ is built from \eqref{thomas precession}.
This can be compared with the Moyal product between plane-waves
\be
e^{i k_1\cdot x} *_M e^{i k_2\cdot x}= e^{i (k_1+ k_2)\cdot x} e^{i \kkk_{12}\cdot \theta} ,
\ee
with $\kkk_{12}\cdot \theta = k_1^{i} k_2^{j}  \theta_\mn$.  There is  a clear analogy, the antisymmetric tensor $\theta_\mn$ being the analogue of the coordinates $j_\mn$. We will show in section \ref{sec:limit DFR}  that this is more than an analogy: the different star products can be related through some limit.

\medskip

We introduce now the Fourier transform of  $\psi(g)$
\be
\hat \psi(x,y)\equiv\int [dg]\, \psi(g)\, e_{g} (x,y)= \int [dk][d\kkk]\, \psi(k,\kkk)\, e_{k,\kkk} (x,y).
\ee
The different field restrictions give then
\bes
\psi(g)\equiv \phi_1(a) &\dr& \hat\psi(x,y)\equiv( \hat \phi_1\circledast\delta_{y=0})(x,y)\label{fourier restriction 1}\\
\psi(g)\equiv \phi_2(a)\delta(h) &\dr& \hat\psi(x,y)\equiv \hat \phi_2(x). \label{fourier restriction 2}
\ees
The field $\phi_1$ depends only on $x$, and the delta function $\delta_{y=0}$ projects $y$ on zero.  The $\circledast $ product is dual to the convolution product on the group. Let $\psi_i\in \cc(\SO(3,1)^+)$, $i=1,2$,  we have
\bes
\int [dg]  e_{g}\, \psi_1\circ \psi_2(g)&=&\int [dg]^3\, e_{g_3}(x,y) \psi_1(g_1)\psi_2(g_2)\delta(g_1g_2g_3\mone)  \nn\\
&=& \int [g]^2\, e_{g_1g_2}(x,y)\, \psi_1(g_1)\psi_2(g_2) \nn\\
&=& \hat\psi_1 \circledast \hat\psi_2(x,y),
\ees
To check that this construction can be related in some ways to Sndyer space-time, let us consider the $\circledast$ product between the coordinates functions.
% They  are obtained in the usual way
%\bes
%x_i= -i\int [dk][d\kkk]\, \delta(k)\delta(\kkk)\,\partial_{k^i} e_{k,\kkk}. \label{x snyder assoc} \\
%y_j= -i\int [dk][d\kkk]\, \delta(k)\delta(\kkk)\,\partial_{\kkk^j} e_{k,\kkk}. \label{y snyder assoc}
%\ees
%We can then identify the star products between the coordinates $x_i$ and $y_j$ and themselves.
To do the calculations explicitly, we choose the embedding coordinates $k_i=p_i=\ka v_i$ and $\kkk_i= \lambda\tan\tilde \eta \,r_i$.
%This choice implies in particular  that
%\be
%\vec \kkk_{12}= \frac{\lambda}{\ka^2}\frac{\vec p_1\wedge\vec p_2}{\gamma_1\gamma_2+ \frac{\vec p_1\cdot \vec p_2}{\ka^2}}.
%\ee
Let us start with $x_i\circledast y_j$.
\bes
x_i\circledast y_j&=& \int [dk]^2[d\kkk]^2\, \delta(k_1)\delta(\kkk_1)\delta(k_2)\delta(\kkk_2)\,\partial_{k_1^i}\partial_{\kkk_2^j}( e_{k_1,\kkk_1}\circledast e_{k_2,\kkk_2})(x,y) \nn\\
&=& \int [dk][d\kkk]\, \delta(k)\delta(\kkk))\,\partial_{k^i}\partial_{\kkk^j}( e^{i k\cdot x} e^{i\mathfrak{k}\cdot y})\nn\\
&=& x_iy_j.
\ees
On the other hand the product $y_j\circledast x_i$ is more interesting. To calculate it we use in particular \eqref{noncom planewave}.
\bes
y_j\circledast x_i&=& \int [dk]^2[d\kkk]^2\, \delta(k_1)\delta(\kkk_1)\delta(k_2)\delta(\kkk_2)\,\partial_{k_2^i}\partial_{\kkk_1^j}( e_{k_1,\kkk_1}\circledast e_{k_2,\kkk_2}) (x,y)\nn\\
&=& y_jx_i-\frac{i}{\lambda}\epsilon_{ij}^kx_k.
\ees
We have therefore that
\be\label{xy sector comm associative}
\com{x_i,y_j}_\circledast= -\frac{i}{\lambda}\epsilon_{ij}^kx_k,
\ee
\ie $y_j$ acts like a rotation on $x_i$. Similar calculations lead to
\bes
&& \com{x_i,x_j}_\circledast=i\frac{\lambda}{\ka^2}\epsilon_{ij}^ky_k,\label{snyder comm associative}, \quad
\com{y_j,y_i}_\circledast= \frac{i}{\lambda}\epsilon_{ji}^ky_k \label{y sector comm associative}
\ees
Instead of using the coordinates $y_i$, we can use the coordinates $j_{mn}$. In this case we have
\bes
&&\com{x_i,x_j}_\circledast=i\frac{\lambda}{\ka^2}j_{ij},\label{snyder comm associative1}\quad \com{x_i,j_{mn}}_\circledast= -\frac{i}{\lambda}( \delta_{ni}x_m - \delta_{mi}x_n), \label{xy sector comm associative1}\\
&& \com{j_{mn},j_{ab}}_\circledast= \frac{i}{\lambda}(\delta_{na} j_{mb}+\delta_{mb} j_{na}-\delta_{nb} j_{ma}-\delta_{ma} j_{nb} ) \label{y sector comm associative1}
\ees
We recognize in \eqref{snyder comm associative1}-\eqref{y sector comm associative1} the Snyder commutation relations since the $y_i$ are identified as rotations from the commutator \eqref{xy sector comm associative}. \emph{We have therefore found an associative star product realization of Snyder space-time non-commutative relations.} It is however important to understand that in this realization   $y_i$ or $j_{mn} $ are \emph{not}  identified with the Lorentz symmetries. These variables encode \textit{extra dimensions}.

\medskip

We can now construct the space-time version of the action \eqref{action field assoc}, by choosing the propagator $\kk(g)$ to be
$\kk(g)= k^2+\kkk^2+m^2$.  By performing the Fourier transform, one obtains
\bes \label{action  associative position}
\ss(\psi)&=& \int [dx][dy]\,\left( (\partial_A\hpsi\circledast \partial^A  \hat\psi)(x,y)+ m^2 (\hat \psi\circledast \hat \psi)(x,y)+\frac{\lambda}{3!}( \hpsi\circledast\hat \psi\circledast \hat \psi)(x,y)\right), \, A=1,..,6.
\ees
considering now the fields restrictions \eqref{fourier restriction 1}, \eqref{fourier restriction 2}, together with the choice of propagator $\kk(g)=\kk(a)= k^2+m^2$, we recover their respective  action
\bes
\ss(\phi_1)&=& \int [dx]\,\left(\delta_0 (\partial_i\hphi_1\circledast  \partial^i  \hat\phi_1 )(x,0)+ m^2 \delta_0(\hat \phi_1\circledast \hat \phi_1)(x,0)+\frac{\lambda}{3!}\delta^2_0( \hphi_1\circledast\hat \phi_1\circledast \hat \phi_1)(x,0)\right), \, i=1,..,3  \label{action with extra delta}\\
\ss(\phi_2)&=& \int [dx][dy]\,\left( (\partial_i\hphi_2\circledast \partial^i  \hat\phi_2)(x)+ m^2 (\hat \phi_2\circledast \hat \phi_2)(x)+\frac{\lambda}{3!}( \hphi_2\circledast\hat \phi_2\circledast \hat \phi_2)(x)\right), \, i=1,..,3 \label{action with extra y}
\ees
The  action $\ss(\phi_1)$ contains product of $\delta_0\equiv \delta(0)$. As we recalled before, in the compact case,  $\delta(0)$ is not divergent, so  the action is well defined. In the non-compact case, the divergences pop up and the action is not well-defined.  In the second action $\ss(\phi_2)$, we are dealing with a field which is not dependent on $y$, but there is nevertheless an integration over $y$, which makes the action badly defined.

We see that we are encountering the same issues as in field theories with extra dimensions where one tries to  project the field theory over a subspace. It is quite interesting to note that  the field restriction $\hpsi(x,y)= \phi_1(x)\delta(y)$ which does not provide a well defined action in the commutative case, can work when dealing  with extra dimensions in the form of a compact Lie algebra.  Unfortunately, the price to pay is the unnatural shape of the momentum conservation in \eqref{action field assoc 1}.

\medskip

We can perform also the Fourier transform for the action defined using the damping function $f$,
\bes \label{action  associative position damped}
\ss^f(\psi)&=& \int [dx][dy]\,\left( (\partial_A\hpsi\circledast \partial^A  \hat\psi \circledast \hat f)(x,y)+ m^2 (\hat \psi\circledast \hat \psi)(x)+\frac{\lambda}{3!}( \hpsi\circledast\hat \psi\circledast \hat \psi \circledast \hat f)(x,y)\right), \, A=1,..,6.
\ees
The  choice $f(g)= \delta(a)$ for $g=ah$ leads to $\hat f(x,y)= \delta(y)$. With this choice in mind,  the previous action becomes
\bes \label{action  associative position damped 1}
\ss^f(\psi)&=& \int [dx]\,\left( (\partial_A\hpsi\circledast \partial^A  \hat\psi \circledast \hat f)(x,0)+ m^2 (\hat \psi\circledast \hat \psi)(x,0)+\frac{\lambda}{3!}( \hpsi\circledast\hat \psi\circledast \hat \psi \circledast \hat f)(x,0)\right), \, A=1,..,6.
\ees
We recovered an action localized on the hyperplane $y=0$. Considering this damping function and the restriction $\hpsi(x,y)=\hat{\phi}_2 (x)$ clearly regularise the action \eqref{action with extra y}. In this case, one can check that the $\circledast$ product is equivalent to the non-associative product $\st$.

\subsection{Deformed Poincar\'e symmetries?}

Since we have defined a dual space-time to the momentum space  $\SO(3,1)^+$, we can now infer what is the realisation of the translation symmetry and check the invariance of the different actions we have constructed.  We consider the usual translation action on coordinates
$$x\dr \tilde x= x+ \varepsilon, \quad y\dr \tilde y= y+ \tilde \varepsilon.$$
The field and the product of fields   transform then in momentum space as
\bes
\hpsi( x, y)\dr  \hpsi(\tilde x,\tilde y) %&\dr& \int [dk][d\kkk]\,   e^{ik\cdot\tilde x}e^{i\kkk\cdot\tilde y} \phi(k, \kkk)\nn\\
&\Rightarrow &\psi(g)\,\dr\,  e_g(\varepsilon,\tilde \varepsilon)\, \psi(g)\\
(\hpsi \circledast \hpsi)( x, y)\dr (\hpsi \circledast \hpsi)(\tilde x,\tilde y) %&=& \int [dk]^2[d\kkk]^2\, e^{i(k_1\oplus k_2')\cdot \tilde x }\,  e^{i (\kkk_{12'}\ominus\kkk_1 \ominus \kkk_2) \cdot \tilde y} \, \phi(k_1,\kkk_1)\phi(k_2,\kkk_2)  \nn\\
&\Rightarrow& \psi(g_1)\phi(g_2)\, \dr \, e_{g_1g_2}(\varepsilon,\tilde \varepsilon )\, \psi(g_1)\phi(g_2).
\ees
The group structure of $\SO(3,1)^+$ encodes the deformation of the translation symmetry $\R^6$, in the momentum space representation. This is the standard construction met for example in the deformation of $\R^3$ translation into $\SU(2)$ \cite{noui} and $\R^4$ into $\AN_3$ \cite{laurent-kowalski}. The action \eqref{action field assoc} is now easily shown to be invariant under the translations symmetry, thanks to the conservation of momenta.

Since we have proposed various field restrictions on the $x$ sector, we check the action of translations on the sector $x$ only. We have then ($g=ah$)
\bes
\hpsi( x, y)\dr  \hpsi(\tilde x,y) &\Rightarrow &\psi(g)\,\dr\,  e_a(\varepsilon)\, \psi(g), \\
(\hpsi \circledast \hpsi)( x, y)\dr (\hpsi \circledast \hpsi)(\tilde x, y)&\Rightarrow& \psi(g_1)\phi(g_2)\, \dr \, e_{(a_1\cdot a_2)h_{12}}(\varepsilon,\varepsilon )\, \psi(g_1)\phi(g_2).
\ees
When considering the transformation of a product of fields on the sector $x$, the non-commutative structure generate also a translation in the $y$ sector. This is quite natural if we recall that the deformation of the translations is such that the non-commutative relation is transformed consistently under translations
\be
[x_i,x_j]_\circledast = i\epsilon_{ij}^k y_k \, \dr \, [ x_i + \varepsilon_i,  x_j + \varepsilon_j]_\circledast = i\epsilon_{ij}^k ( y_k+ \varepsilon_k).\ee
This has however a direct consequence for the field restrictions we have introduced: they do break the translational symmetry in the $x$ sector. The introduction of the damping function has the same effect, if it is not transforming as a field. The actions \eqref{action with extra delta}, \eqref{action with extra y} do not possess the translational symmetry inherited from the main theory \eqref{action associative position}. In the commutative case, the damping function $\hat f(x,y)=\delta(y)$ would lead to theory with translation symmetry on the hyperplane $(x,0)$. The non-commutative structure spoils this translational  symmetry.

The action of the group $\SO(3,1)^+$ on $\R^6$ is recovered from the Fourier transform: this is  the adjoint action. The infinitesimal generators $(K_i,J_i)$ act as follows
\bes
\com{K_i,x_j}=- i\epsilon_{ij}^k y_k\quad \com{K_i,y_j}= -i \epsilon_{ij}^kx_k\\
\com{J_i,x_j}= i\epsilon_{ij}^k x_k \quad \com{J_i,y_j}= i\epsilon_{ij}^k y_k.
\ees
This means in particular that  $\SO(3)$ acts on the dual coordinates $(x_i,y_j)$ by the adjoint action.  It is  clear that we are not dealing with a deformation of the Poincar\'e group acting on $\R^6$, since in this case the Lorentz group should be 15 dimensional and not 6 dimensional as   $\SO(3,1)^+$.

\subsection{Relation to the  DFR and Moyal space-times}\label{sec:limit DFR}
Following our construction on the Sndyer space, we present the construction of a field theory on the Doplicher, Fredenhagen and Roberts non-commutative space (DFR space) \cite{dfr}. We recover many of the results of    \cite{carlson}. The DFR space  is encoded in the coordinates operators $(X_i,\Theta_{ij})$,  where $\Theta_{ij}$ is an antisymmetric operator with dimension length ($\hbar=c=1$). The construction was initially done in 4d Lorentzian, but it applies also in the 3d Euclidian case, which we consider from now on.  These operators satisfy the following commutation relations, (with $\ell$ a constant with dimension length)
\be\label{DFR}
\com{X_i,X_j}= i \ell\Theta_{ij}, \quad \com{X_i,\Theta_{kl}}=0=\com{\Theta_{mn},\Theta_{ij}}, \quad i,j,m,n=1,2,3,
\ee
which makes the set $(X_i,\Theta_{ij})$ a Lie algebra $\aa$ since the Jacobi identities are satisfied. The $\Theta$ sector forms an abelian sub Lie algebra $\mathfrak{h}\sim \R^3$ of $\aa$. One of the key features of this space is that it is covariant under the Lorentz transformations acting on the $X_i$ sector. Explicitly, if $X_i$ transforms as a vector, the variables $\Theta_{ij}$ transform as a tensor under Lorentz transformations given by $\SO(3)$.   The commutators \eqref{DFR} are then also transforming covariantly under $\SO(3)$. Another important feature is that the translations do act on the $X_i$ as in the usual commutative case, which means in particular that the $\Theta_{ij}$ is actually invariant under the translations applied on the $X_i$ sector \cite{dfr}.

We consider now the group $\ggg$ generated from the Lie algebra $\aa$. We can write an element $g\in \ggg$ using the coordinate parameters $(\eta, \tilde \eta, \vec \oll, \vec \qqq)$
\be
g= bh=e^{i  \eta \oll \cdot X}e^{i \tilde\eta\qqq\cdot \Theta }, \quad \qqq \cdot \Theta\equiv \qqq^{ij}\Theta_{ij}.\ee
The product of group elements generate a modified addition for the coordinate parameters
\bes \label{product DFR moyal}
&& b_1h_1 b_2h_2=( b_1\cdot  b_2 )h_{12}h_1h_2\\
&& e^{i \eta_1\vec \oll_1\cdot X}e^{i\tilde \eta_1\qqq_1 \cdot \Theta}\, e^{i \eta_2\vec \oll_2\cdot X}e^{i\tilde\eta_2\qqq_2 \cdot \Theta}=e^{i (\eta_1\vec \oll_1+\eta_2 \vec \oll_2)\cdot X}\, e^{\frac{i\ell}{2} \eta_1\eta_2\oll_1\cdot \Theta \cdot \oll_2}\, e^{i(\tilde \eta_1\qqq_1+\tilde \eta_2\qqq_2)\cdot \Theta}, \,\, \oll_1\cdot \Theta \cdot \oll_2 \equiv  \oll_1^i \oll_2^j \Theta_{ij}.
\ees
Momentum variables $(q,\lll)$ are constructed from the coordinate parameters
\bes
\vec q= \frac{1}{\ell}\tilde f_1(\eta)\vec \oll,  \quad
\vec \lll=\frac{1}{\ell} \tilde f_2 (\tilde\eta)\vec \qqq.
\ees
According to the choice of $\tilde f_1$, $\tilde f_2$, the product \eqref{product DFR moyal} induces a modified sum for the momenta. Making the specific choice $\tilde f_1(\eta)=\eta$ and $\tilde f_2(\tilde \eta)=\tilde \eta$, we obtain the sum
\bes
\vec q_1\oplus \vec q_2& =& \vec q_1 + \vec q_2\\
(\lll_1\oplus  \lll_2 )_{ij}&=& \frac{\ell}{2}q_{1i}  q_{2j}+\lll_{1ij}+\lll_{2ij}
\ees
The momentum addition on the $q$ sector is the standard one, whereas there is a non-trivial sum on the $\lll$ sector. Considering the measures on flat spaces, respectively $[db]\sim d^3q$ and $[dh]\sim d^3\lll$, which are the natural measures for this group (as we shall see later), we construct an action for the scalar field  $\psi(g)=\psi(b,h)= \psi(\vec q, \vec \lll)$
\be
\ss_{dfr}(\psi)= \int [dg]^2\phi(g_1)\kk(g_1)\phi(g_2)\delta(g_1g_2)+\frac{\lambda}{3!}\int  [dg]^3\phi(g_1)\phi(g_2)\phi(g_3)\delta(g_1g_2g_3).
\ee
In \cite{carlson}, the authors considered the action constructed with the help of a damping function $f(g)= \delta(b)f(h)$
\be
\ss_{dfr}^f(\psi)= \int [dg]^2[dh]\phi(g_1)\kk(g_1)\phi(g_2)f(h_3)\delta(g_1g_2h_3)+\frac{\lambda}{3!}\int  [dg]^3[dh]\phi(g_1)\phi(g_2)\phi(g_3)f(h_4)\delta(g_1g_2g_3h_4).
\ee
We now construct the space-time representation of this action, by introducing the Fourier transform in the usual way. First we considers the   coordinates functions $(x_i,\theta_{ij})\in \R^6$, which are the analogue of the operators   $(X_i,\Theta_{ij})$. Given a choice of momentum coordinates $(q,\lll)$, we introduce    the plane-wave  $ e_g(x,\theta)\equiv e^{i (q\cdot x+\lll \cdot \theta)}$. The product of plan-waves is non-trivial in order to encode the non-trivial sum \eqref{product DFR moyal}.
\bes
(e_{g_1}*_{dfr} e_{g_2})(x,\theta)&\equiv& e_{g_1g_2}(x,\theta)\\
 e^{i (q_1\cdot x+\lll_1 \cdot \theta)}\, *_{dfr}  \, e^{i (q_2\cdot x+\lll_2 \cdot \theta)}&\equiv&  e^{i ((q_1+q_2)\cdot x+(\lll_1+\lll_2) \cdot \theta+\frac{i\ell}{2} q_1\cdot \theta\cdot q_2)}.\nn
\ees
We can then introduce the Fourier transform of a given distribution $\psi(g)$,  considering the measure $ [dg]=[d^3q][d^3\lll]$ on the group $\ggg$.
\be
\hat \psi (x,\theta)= \int [dg] \, e_g(x,\theta) \psi(g).
\ee
The different commutators of the coordinates $(x,\theta)$  are calculated using the Fourier transform and one obtains the star product realization of the DFR commutations relations.
\bes
&&\com{x_i,x_j}_{*_{dfr}}= i \ell\theta_{ij}, \quad \com{x_i,\theta_{kl}}_{*_{dfr}}=0=\com{\theta_{mn},\theta_{ij}}_{*_{dfr}}, \quad i,j,m,n=1,2,3,
\ees
Furthermore, one can give an explicit realization of the star product \cite{carlson}
\be
(\hpsi_1*_{dfr}\hpsi_2)(x,\theta)= \psi_1(x,\theta)e^{\frac{i}{2}\ell\overleftarrow{\partial_i} \theta_{ij}\overrightarrow{\partial_j}}\psi_2(x,\theta)
\ee
In particular, we recover the well-known property of the Moyal product when considering the integral of the  star product of two distributions
\be
\int [dx][d\theta]\, (\hpsi_1*_{dfr}\hpsi_2)= \int [dx][d\theta]\, \hpsi_1\hpsi_2.
\ee
This star product is therefore very close to the Moyal product. The main difference  is that in the Moyal case, the tensor $\theta_{ij}$ is not transforming under the Lorentz transformations, as it does from the DFR perspective. The Moyal star product can nevertheless be retrieved from the DFR star product by simply \textit{fixing} $\theta_{ij}=\ov\theta_{ij}$ to an arbitrary value, such that it does not transform anymore under the Lorentz transformations. We deal therefore with a Moyal non-commutative space-time that breaks the Lorentz symmetry (as opposed to the Moyal space-time where the Lorentz symmetries are deformed).

As usual, the convolution product is the dual of the star product and straightforward calculations now give  the space-time representation of the actions, with the choice of propagator $\kk(g)= k^2+\kkk^2+m^2$
\bes \label{action   position dfr 1}
\ss(\psi)&=& \int [dx][d\theta]\,\left( (\partial_A\hpsi*_{dfr} \partial^A  \hat\psi)+ m^2 (\hat \psi*_{dfr} \hat \psi)+\frac{\lambda}{3!}( \hpsi*_{dfr}\hat \psi*_{dfr} \hat \psi)\right), \, A=1,..,6.\\
\ss^f(\psi)&=& \int [dx][d\theta]\,\left( (\partial_A\hpsi*_{dfr} \partial^A  \hat\psi*_{dfr}\hat f)+ m^2 (\hat \psi*_{dfr} \hat \psi*_{dfr}\hat f)+\frac{\lambda}{3!}( \hpsi^{*_{dfr}3}*_{dfr}\hat f)\right), \, A=1,..,6. \label{action   position dfr 2}
\ees
The last action is the action considered in \cite{carlson}, with extra assumptions on the damping  (weighting) function $\hat f(x,\theta)\equiv W(\theta)$. They also considered the field restriction  $\hpsi(x,\theta)\equiv \hphi_2(x)$, that is a field depending on $x$ only.
\bes \label{action   position dfr 3}
\ss^f(\psi)&=& \int [dx][d\theta]\,\left( (\partial_i\hphi_2*_{dfr} \partial^i  \hphi_2*_{dfr}\hat W)+ m^2 (\hphi_2*_{dfr} \hphi_2*_{dfr}\hat W)+\frac{\lambda}{3!}( \hphi_2^{*_{dfr}3}*_{dfr}\hat W)\right).
\ees
In this case, it is easy to check that this action is invariant under the translations, if there are the realized in the DFR way, that is the coordinates transform in the usual way and the coordinates $\theta$ are \textit{invariant} under such transformation.
\bes\label{dfr translation}
&&x_i\dr x_i+\varepsilon_i, \, \theta_{ij}\dr \theta_{ij}, \quad \com{x_i+\varepsilon_i ,x_j+\varepsilon_j }_{*_{dfr} }=\theta_{ij}.
\ees

\medskip

The DFR construction shares many similarities with the construction described in the context of Snyder space-time in the previous section. The two non-commutative spaces  can in fact be  related through a limit\footnote{In \cite{carlson}, we have $b=\frac{1}{\lambda},$ $a=\frac{1}{\ka}$. }  \cite{carlson}.
\be\label{limit}
\frac{\lambda}{\ka^2}\,\dr\, \ell \textrm{ with }\left(\begin{array}{l} \ka\,\dr\, \infty \\ \lambda\,\dr\, \infty\end{array}\right.,
\ee
With this limit, the star product commutators \eqref{y sector comm associative1} and  \eqref{xy sector comm associative1} become
\bes
\com{x_i,x_j}_\circledast&=&i\frac{\lambda}{\ka^2}j_{ij}\, \dr\, i\ell\,j_{ij},\label{DFR comm associative1}\\
\com{x_i,j_{mn}}_\circledast&=& -\frac{i}{\lambda}( \delta_{ni}x_m - \delta_{mi}x_n)\,\dr\, 0, \label{DFR xy sector comm associative1}\\
\com{j_{mn},j_{ab}}_\circledast&=& \frac{i}{\lambda}(\delta_{na} j_{mb}+\delta_{mb} j_{na}-\delta_{nb} j_{ma}-\delta_{ma} j_{nb} )\,\dr\, 0 \label{DFR y sector comm associative1}
\ees
We can therefore identify $j_{mn}$ with $\theta_{mn}$.

The limit provides also to recover  the DFR addition law for momenta, provided we choose  consistent definitions for the momenta $(\vec k, \vec \kkk)$ and $(\vec q, \vec \lll)$. We demand  that $\vec k\dr \vec q$ and $\vec \kkk\dr\vec \lll$  and
\bes
&&\vec \kkk_{12}\,\dr\, \frac{\ell}{2} \vec q_1\wedge\vec q_2,\\
&& \vec k_1\oplus\vec k_2\,\dr\,\vec q_1+\vec q_2\\
&&e^{i\kkk_{12}\cdot y}\,\dr\, e^{i\frac{\ell}{2}\, q_1^mq_2^nj_{mn}}\\
&&e^{ik_1\cdot x}\circledast e^{ik_2\cdot x}\,\dr \, e^{i((q_1+q_2)\cdot x+\frac{\ell}{2}(q_1^mq_2^nj_{mn}))}
\ees
As an  example, the  choices for the Sndyer case $f_1(\eta)=\sinh\eta$ or $f_1(\eta)=\tanh\eta$ with the choice $ f_2(\tilde \eta)=\tan\tilde \eta $ are consistent  with  the  coordinate choice in the DFR case  $\tilde f_1(\eta)=\eta$ and $ \tilde f_2(\tilde \eta)=\tilde \eta$.

This limit enables to see that the measure on $\ggg$ is given by the measure on $\R^6$. Indeed starting from the measure $[dg]$ on $\SO(3,1)^+$ and considering the limit with the consistent choice of momenta as above, it is easy to see that    $[dg]\dr d^3qd^3\lll$, the  measure on $\ggg$.

Before concluding this section, we would like to comment on the symmetries. The choice of propagator and field restriction makes clear that the Lorentz symmetry is simply $\SO(3)$. There is however a subtlety with the translation invariance. Indeed, the translation symmetry constructed from the Snyder perspective under the limit \eqref{limit} makes sure that the commutator $[x_i,x_j]_{*_{dfr}}= i\ell\theta_{ij}$ is transforming covariantly, that is
\be
[x_i,x_j]_{*_{dfr}}= i\ell\theta_{ij} \dr [x_i+\varepsilon_i,x_j+\varepsilon_j]_{*_{dfr}}= i\ell(\theta_{ij}+\epsilon_{ij}^k\varepsilon_k).
\ee
This translation action is not the DFR type translation action \eqref{dfr translation}.
The action \eqref{action   position dfr 3} constructed from the field restriction and the damping function is not invariant under the translation symmetry inherited from the Sndyer approach, since the field restriction and the damping function clearly break this symmetry. It is not clear to us if the DFR realization of the translations can be related to  through a different realization of the translation symmetry in the Snyder case, for example due to the existence of another sort of differential calculus. We leave this question for further  investigations.

%%%%%%%%%%%%%%%%%%%%%%%%%%%%%%%%%%%%%%%%%%%%%%%%%%%%%%%%%%%%%%%%%
\section{Some physical comments}\label{sec:physics assoc}

\subsection{Different approximation schemes}

We have considered two approaches  in the previous sections to construct  different types of action on non-commutative spaces which satisfied Snyder's commutations relations. They can be  generalized to any dimension and in particular to the 4d Sndyer case with $G=\SO(4,1)$ and $H=\SO(3,1)$. These approaches are very different, since in one case there is a non-associative structure, and in the second the fundamental structure is associative. We showed how in the latter case, one could recover the former case, using the damping function $\hat f(x,y)=\delta (y)$ and a field restriction $\hpsi(x,y)=\hphi_2(x)$.

This has a  natural physical interpretation: starting with a field theory  with extra degrees of freedom, introduced through extra-dimensions, we intend to describe an effective theory where we want to damp  the impact of these extra dimensions on the relevant space-time dimensions (specified by the $x$). The actions with the damping term can be simply rewritten as
\be
\ss^f(\hpsi)=\int [dx][dy]\, (\Ll(\hpsi, \partial\hpsi) *\hat f)(x,y),
\ee
and in the DFR case, this takes the even more suggestive shape
\be
\ss^f(\hpsi)=\int [dx][dy]\,\hat f(y) \Ll(\hpsi,  \partial\hpsi).
\ee
The damping function $\hat f$ consists in a modification of the trace which precisely truncates these extra degrees of freedom.  To achieve this,  the function is therefore assumed to be positive,  for any large $|y_{i}|$ , it goes to zero quickly enough and it is even \ie
\be
\int [dy]\, \hat W(y) y_{{i}}=0.
\ee
Unlike the commutative case, if we consider the field restriction $\hpsi(x,y)=\hphi_2(x)$, the field theory deduced from the fundamental theory still feels the extra dimensions, which are hidden in the star product, \ie the non-commutative structure. By considering the extreme case where the damping function localizes on the $y=0$ hyperplane, \ie $\hat W(y)=\delta(y)$, we  recovered the non-associative approach. In this sense the non-associative field theory  appears as a specific approximation of the associative field theory constructed on a space-time with  extra dimensions.

We can interpret physically this procedure by considering the associative field theory constructed on the group manifold as the fundamental field theory, with the the $x$-sector being our usual physical sector and the $y$-sector representing extra degrees of freedom. Integrating over the $y$ sector would then amount to tracing out these degrees of freedom, which would lead to an apparent violation of momentum conservation in the $x$-sector (which happens in our framework) and to potential non-associative effects. From this point of view, localizing on the $y=0$ hyperplane could then be interpreted as working on a (degenerate) ground state of the system as a first order approximation to the theory. Then high energy dynamics would require investigating the effect of fluctuations out of the $y=0$ hyperplane and we would be able to probe the whole structure of the full associative field theory. We underline that this is simply a proposal for a physical  interpretation, which would require further investigation. 

Nevertheless, we would like to compare this situation to a (scalar) field theory on the 4d Lorentzian Snyder space-time and its possible relation to 4d quantum gravity. In \cite{dsr-group}, a derivation of effective non-commutative field theory for matter fields from group field theories for spinfoam models of 4d quantum gravity was presented. These group field theories are inspired from the reformulation of general relativity as a BF theory with potential based on the $\SO(4,1)$ gauge group. Starting from there, we naturally obtain effective associative field theory with $\SO(4,1)$ as momentum space. In this context, we face the same issues as presently when we need to either trace out the $\SO(3,1)$ sector or localize it in order to obtain a field theory on the Snyder spacetime dual to the coset momentum space $\SO(4,1)/\SO(3,1)$. We then have to address the physical relevance of the $\SO(3,1)$-sector of the field theory, which is actually the issue left unanswered in the approach presented in \cite{dsr-group}.

\medskip

In a similar way the DFR field theory appears as an approximation of the fundamental associative field theory. A more general damping function than  $\hat W(y)=\delta(y)$ is allowed and we have also to perform the limit $\ka, \lambda\dr \infty$ with $\frac{\lambda}{\ka^2}=\ell$ fixed. This limits amounts to abelianize the extra-dimensions, while keeping some non-commutativity. This is a semi-classical limit since we send the Planck mass $\ka$ to $\infty$. The non-commutativity is retained in this regime by sending the other mass scale $\lambda$ to $\infty$ too. This is another approximation of the fundamental field theory defined over space-time embedded in extra dimensions.

%\begin{tabular}{rcl}
%& Associative field theory on $\SO(4,1)$ & \\
%& $ \swarrow \quad \searrow$ &  \\
%DFR space && Non-associative scalar field theory on $\SO(4,1)/\SO(3,1)$
%\end{tabular}

\subsection{Momentum and multi-particles states}
In both approaches, associative or not, we are dealing with a curved momentum space just as in the context of $\ka$-Minkowski. In this context, many issues are well-known \cite{kowalski}. For example, there are different possible choices of momentum. In the $\ka$-Minkowski case, one can "solve" this issue by analyzing the differential calculus: it provides a natural recipe to choose what we call momentum. We do not have such calculus at this time in the Snyder space-time, so we are left with some ambiguities. We have been dealing with the canonical choices, the Snyder and embedding coordinates, in analogy with $\ka$-Minkowski. Snyder's momentum $P_{\mu}$ is  analog to the choice of momentum made in the bicrossproduct construction of $\ka$-Minkowski, which imposes a bound on energy \cite{jerzy-giovanni}. Note that the Snyder's choice for momentum  and the embedding coordinates choice $p_\mu$ seem to be qualitatively different, since in the first case, there is a bound on the rest mass, $P_\mu P^\mu= m^2\leq \ka^2$ whereas in the latter there is no obvious bound on momentum. The choice of momentum is of course very important to relate the theory to experiments. It also affects by construction the $\st$ product one uses, since this product is defined so that  the product of  plane-wave encodes the momenta addition.

This bound on momentum (on energy or rest-mass) is problematic when considering the sum of momenta. Indeed, by construction the resulting momentum will  be bounded in the same way. This is the so-called "soccer ball" problem. It is then unclear how to recover a system with a momentum that do not respect this bound, that is how to recover a "classical" object. Using the Snyder momentum coordinates, we do not escape this problem. At the contrary, again, it seems that by using the embedding coordinates $p_\mu$ we do not have this issue.

$\ka$-Minkowski physics is also characterized by a non-commutative addition of the momenta. From the first approach we presented (section \ref{sec:field on coset}), we see that in Snyder's space-time  the addition is non-commutative \emph{and non-associative}. It has been shown in the context of $\ka$-Minkowski that a careful analysis of the partition function for the scalar field theory with a source term, naturally generates all the possible types of interactions \cite{arzano}. More explicitly, one obtains  a symmetrization of the momenta conservation. In the case of a $\hphi^3$ interaction, we would get for example
\be
\delta(k_1\oplus k_2\oplus k_3)\dr \sum_\sigma \delta(k_{\sigma(1)}\oplus k_{\sigma(3)}\oplus k_{\sigma(3)}),
\ee
where $\sigma$ is a permutation. In this sense, even though one conservation rule is not symmetric under the permutation of momentum, we have many conservation laws that imply that the total amplitude is symmetric under particle exchanges. Such approach could be extended to consider all the possible grouping of products in order to have at the end a symmetric amplitude. In the case of a $\hphi^3$ interaction, we would get for example
\be
 \sum_\sigma \delta(k_{\sigma(1)}\oplus (k_{\sigma(3)}\oplus k_{\sigma(3)}))\dr \sum_\sigma \delta(k_{\sigma(1)}\oplus (k_{\sigma(3)}\oplus k_{\sigma(3)}))+ \sum_\sigma \delta((k_{\sigma(1)}\oplus k_{\sigma(3)})\oplus k_{\sigma(3)}).
\ee

\bigskip

In the second approach (section \ref{sec:field on group}),  we do have  an associative momenta addition, though still non-commutative. This simplification is however at the cost of introducing the extra momentum sector $\kkk$ which physical interpretation is open. In the context of the DFR model, this sector is interpreted as some version of the (fuzzy) Kaluza-Klein approach, where one deals with extra dimensions. A similar interpretation can hold in the Snyder context.

\section*{Discussion}
We have presented two approaches to describe a scalar field theory on a non-commutative space-time that reproduces the Snyder's commutation relations. We have in particular presented two star product realizations of this non-commutative space. The two approaches are very distinct, since in one case the star product is non-associative, whereas in the second one it is associative. This difference comes from a different initial set up: in the non-associative case, we considered  the field theory living only on the upper hyperboloid interpreted as momentum space, whereas in the associative case, momentum space is the group $\SO(3,1)^+$, in which the upper hyperboloid is naturally embedded. In the former case, we identified the deformed Poincar\'e symmetries which are consistent with this non-commutative structure. This deformation has not been identified in the classification of the deformation of the Poincar\'e symmetries \cite{zak}, since this classification looked at the deformations which preserved the (co-)associativity. The construction of the new quantum group encoding this deformation is currently under investigations \cite{flo}. One of the key-questions to answer in  the curved momentum space setting is the analysis of the Noether charges. This analysis can be done in the chosen star product realization but a complete analysis would rely on classification of the differential calculus on this non-commutative space. We refer to \cite{flo} for further details.

To consider the full group as momentum space  can be seen as constructing a scalar field theory on a non-commutative  space-time with extra dimensions, and we presented different ways to reduce the theory on relevant space-time.  We showed moreover how the DFR non-commutative space-time is naturally related to this construction through some limit. We also showed how the non-associative model can be related to this  extra-dimensions model if one considers some damping function and field restriction. In fact both the DFR space and the non-associative space can be seen as specific approximations of the extra-dimension model. We pinpointed that there exists a priori a mismatch between the translation symmetry obtained by considering  the DFR limit in the extra-dimensions model and the DFR realization. It is unclear to us if there the two realizations could be related or if in fact the DFR translation realization is an different realization which can not be traced back to the extended model, and simply due to the abelian features of the DFR space. To answer this question, one should consider the classification of the differential calculus on $\sll(2,\C)$ and the possible deformations of the Poincar\'e symmetries acting on $\R^6$.

We have presented the construction in a 3d Euclidian setting, which was a nice simple laboratory to present all the features of the two different approaches. Our calculations can be extended to the 4d Lorentzian case, the calculations becoming just cumbersome. To define a scalar field theory on the Lorentzian Sndyer's space-time, one should  consider the group $\SO(4,1)$ and the subgroup $\SO(3,1)$, so that momentum space is the de Sitter space $\SO(4,1)/\SO(3,1)$. This space is embedded in $\R^5$ through $\lbrace v^A\in \R^5, \, v^Av_A=-\ka^2\rbrace$ and  we have two open sets to  cover it. In the Sndyer coordinates $P_\mu= \ka \frac{v_\mu}{v_4}$,  these two sets are simply given in terms of $v_4>0$ and $v_4<0$.  Just as in the case of the hyperboloid, one should work only in one set, choosing for example $v_4>0$. This restriction is consistent with the action of the  Lorentz group $\SO(3,1)$, so that there will be a natural realization of these symmetries, just as the $\SO(3)$ symmetry was realized in our example. Once again one can discuss a field theory on the  full group $\SO(4,1)$ and implement the damping function and field restriction to recover the non-associative construction and the DFR space. This analysis would then motivate the studies of deformation of the Poincar\'e group acting on $\R^{10}$ and the analysis of the differential calculus on the associated non-commutative space. We leave these ideas for further investigations.

To conclude we have defined two approaches to define in a natural way a scalar field classical action living on Snyder'space-time, with the relevant Poincar\'e symmetries. Snyder's initial motivation was to introduce this non-commutative space to regularize the UV divergences in field theory. We have now all the tools to construct  the quantum field theory for this space and  to perform the divergences analysis, to see at last if Snyder's guess was right or not.

\section*{Acknowledgement}
FG would like to thank S. Doplicher for pointing out the reference \cite{carlson}.

%%%%%%%%%%%%%%%%%%%%%%%%%%%%%%%%%%%%%%%%%%%%%%%%%%%

\end{document}